\documentclass[useAMS,usenatbib]{mn2e}
\usepackage{times}  % to get nice fonts for PDF
\usepackage{listings}
\usepackage{graphics}
\usepackage{subfigure}
\usepackage{graphicx}
\usepackage{graphicx,xspace}
\usepackage{amssymb}
\usepackage{amsmath}
\usepackage{hyperref}
\usepackage{aas_macros}
\usepackage{lscape}
\usepackage{rotating}
\usepackage{placeins}
\usepackage{natbib}
\usepackage{color}
\usepackage{caption}
\usepackage{float}

\newcommand{\powmes}{{\sc{powmes}}\xspace}
\newcommand{\python}{{\sc{python}}\xspace}
\newcommand{\subfind}{{\sc{subfind}}\xspace}
\newcommand{\nbodykit}{{\sc{nbodykit}}\xspace}
\newcommand{\NGenIC}{{\sc{N-GenIC}}\xspace}
\newcommand{\CAMB}{{\sc{CAMB}}\xspace}
\newcommand{\gadget}{{\sc{gadget-3}}\xspace}
\newcommand{\hM}[1]{$10^{#1}\, \mathrm{M_\odot}/h$}
\newcommand{\M}{$M_{500,\mathrm{crit}}$}
\newcommand{\radc}{$r_{500,\mathrm{c}}\,$}
\newcommand{\radm}{$r_{200,\mathrm{m}}\,$}
\newcommand{\ksc}[2]{$k\,{#1}\,{#2}\,h\mathrm{\,Mpc^{-1}}$}

\newcommand{\Pmm}{P_\mathrm{mm}}

\newcommand{\Pmhi}{P_{\mathrm{mh},i,\Delta}}
\newcommand{\Pmnh}{P_{\mathrm{mnh},\Delta}}
\newcommand{\Pmhtwo}{P_{\mathrm{mh,200m}}}
\newcommand{\Pmhitwo}{P_{\mathrm{mh},i,\mathrm{200m}}}
\newcommand{\Pmhifive}{P_{\mathrm{mh},i,\mathrm{500c}}}
\newcommand{\Pmnhtwo}{P_{\mathrm{mnh,200m}}}
\newcommand{\Pmmp}{P'_\mathrm{mm}}
\newcommand{\Pmhp}{P'_{\mathrm{mh},\Delta}}

\newcommand{\Pmnhp}{P'_{\mathrm{mnh},\Delta}}
\newcommand{\Pmmpp}{P''_\mathrm{mm}}
\newcommand{\Pmhpp}{P''_{\mathrm{mh},\Delta}}

\newcommand{\Pmnhpp}{P''_{\mathrm{mnh},\Delta}}
\newcommand{\Pmhitwop}{P'_{\mathrm{mh},i,\mathrm{200m}}}
\newcommand{\Pmhifivep}{P'_{\mathrm{mh},i,\mathrm{500c}}}
\newcommand{\Pmhitwopp}{P''_{\mathrm{mh},i,\mathrm{200m}}}
\newcommand{\PmAi}{P_{\mathrm{mA},i}}
\newcommand{\PmAip}{P'_{\mathrm{mA},i}}
\newcommand{\bM}{b_{i,\Delta}}
\newcommand{\fM}{f_{\mathrm{M},i,\Delta}}
\newcommand{\fMitwo}{f_{\mathrm{M},i,\mathrm{200m}}}
\newcommand{\fMifive}{f_{\mathrm{M},i,\mathrm{500c}}}
\newcommand{\fMp}{f'_{\mathrm{M},i,\Delta}}
\newcommand{\fret}{f_{\mathrm{ret},i,\Delta}}
\newcommand{\frettwo}{f_{\mathrm{ret},i,\mathrm{200m}}}
\newcommand{\fretfive}{f_{\mathrm{ret},i,\mathrm{500c}}}
\newcommand{\fbi}{\bar{f}_{\mathrm{b},i,\Delta}}
\newcommand{\fbci}{f_{\mathrm{bc},i,\Delta}}

\setcitestyle{notesep={ },round,aysep={},yysep={}}
\title[Halo contributions to the power spectrum]
{The contribution of massive haloes to the matter power spectrum in the presence of AGN feedback}
\author[M. L. van Loon \& M. P. van Daalen]{
\newauthor M. L. van Loon\thanks{E-mail: mvanloon98@gmail.com}$^{1}$,
Marcel P. van Daalen$^{1}$
\\
$^{1}$Leiden Observatory, Leiden University, P.O. Box 9513, 2300 RA Leiden, the Netherlands\\
}
\begin{document}
\date{\today}
\pagerange{\pageref{firstpage}--\pageref{lastpage}} \pubyear{2023}
\maketitle
\label{firstpage}

\begin{abstract} %currently 248 words (if you count math expressions as 1 word), MNRAS limit is 250
The clustering of matter, as measured by the matter power spectrum, informs us about dark matter and cosmology, as well as baryonic effects on the distribution of matter in the universe.
Using cosmological hydrodynamical simulations from the cosmo-OWLS and BAHAMAS simulation projects, we investigate the contribution of power in haloes with various masses, defined by particles within some overdensity region, to the full power spectrum, as well as the power ratio between baryonic and dark matter only (DMO) simulations for a matched (between simulations) and an unmatched set of haloes.
We find that the presence of AGN feedback suppresses the power on all scales for haloes of all masses examined ($10^{11.25}\leq$\M$\leq$\hM{14.75}), by ejecting matter from within \radc to \radm and potentially beyond in massive haloes (\M$\gtrsim$\hM{13}), and likely impeding the growth of lower-mass haloes as a consequence.
A lower AGN feedback temperature drastically changes the behaviour of high-mass haloes (\M$\geq$\hM{13.25}), damping the effects of AGN feedback at small scales, \ksc{\gtrsim}{4}.
For \ksc{\lesssim}{3}, group-sized haloes (\hM{14\pm0.25}) dominate the power spectrum, while on smaller scales the combined contributions of lower-mass haloes to the full power spectrum rise above that of the group-sized haloes.
%However, the combined contributions of haloes with masses $\geq$\hM{11.25} to the full spectrum does not amount to within $1\%$ of the full power spectrum on any scale, indicating that the power in non-halo particles and cross-power between halo and non-halo particles may have a significant role to play.
Finally, we present a model for the power suppression due to feedback, which combines observed mean halo baryon fractions with halo mass fractions and halo-matter cross-spectra extracted from dark matter only simulations to predict the power suppression to percent-level accuracy down to \ksc{\approx}{10} without any free parameters.
\end{abstract}


\section{Introduction}
\label{intro}
Understanding the way matter clusters is an integral piece in furthering our knowledge about the Universe and its inner workings.
The clustering of matter governs halo collapse and galaxy formation, along with mergers and other processes.
All these processes are heavily dependent on the cosmological parameters, the fundamental parameters describing our universe.
These parameters dictate, among other things, the matter density of the universe, its expansion, and, by extension, how much of the matter is formed into haloes and galaxies, along with their formation timescales.

Although the $\Lambda$CDM model is the most widely established cosmological model, incorporating cold dark matter and dark energy, its parameters are not set.
Observational data from the two largest cosmological probes, the Wilkinson Microwave Anisotropy Probe (WMAP) \cite[see][for the 7- and 9-year results, respectively]{Komatsu_2011,Hinshaw_2013} and Planck \cite[see][for the 2013 and 2015 releases, respectively]{2014planck,planck2015}, provide measurements of the cosmic microwave background (CMB) from which the cosmology of our Universe can be derived, but there is growing evidence that there is tension between these early-Universe probes and late-Universe measurements \citep[see][for a recent review]{Abdalla2022tensionreview}. The latter type of measurements typically study some aspect of the clustering of matter, for example through weak gravitational lensing \citep[e.g.][]{Heymans2021KiDS,Abbott2022DES}, and compare this to theoretical models to derive cosmology.
One particular measure of clustering is the matter power spectrum, generally given as $P(k)$ (or $\Delta^2(k)$): the amplitude squared of matter overdensities as a function of Fourier scale, $k$, which is given by $2\pi/\lambda$, where $\lambda$ refers to the physical scale.
%It can be seen as the amplitude squared of the Fourier transform of the matter overdensity field.

On the largest scales, linear perturbation theory is sufficient to model the power spectrum, originally proposed by \cite{press1974formation} and extended by \cite{bond1991excursion}. 
However, on smaller, non-linear scales, governing halo collapse and galaxy formation, linear theory underestimates the power and additional physics needs to be taken into account.
A widely accepted model incorporating non-linear scales is the analytical halo model \cite[e.g.][]{peacock2000halo, seljak2000analytic, cooray2002halo}, which has been added to by many others \cite[e.g.][]{tinker2008toward, duffy2008dark, semboloni2013effect, mead2015accurate, debackere2020impact, 2021Mead}.
As future weak lensing surveys, enabling high precision ($<1\%$) measurements of (matter) power spectra, draw near, the importance of theoretical understanding and predictions on these non-linear scales increases \cite[e.g.][]{huterer2005calibrating, laureijs2009euclid, 2012Hearin}.

Currently, the precise effects of baryons, e.g. through galaxy formation and evolution, on the matter power spectrum (hereafter referred to as: power spectrum) are still largely undetermined \cite[see][for a review]{chisari2019modelling}.
Countless complex and coupled processes underlie the formation and evolution of galaxies that often play out on small, not directly observable, scales.
Hydrodynamical, cosmological simulations, which can produce a statistically interesting sample of realistic galaxies, offer insights and help to determine which theoretical models best describe observations or which observations are predicted based on theory.
The trade-off between resolution and sample size remains a challenge, although increased computational efficiency and so-called 'subgrid models', which allow unresolved physical processes to be taken into account, have come a long way in the last few decades \citep[for a review, see][]{Vogelsberger2020hydroreview}. %\cite[e.g.][]{2010schaye, Schaye15, McCarthy2017}.

The widely accepted idea that the clustering of all matter is fully determined by that of dark matter was refuted when \cite{2011vanDaalen} found that the presence of AGN feedback has a profound impact on the matter power spectrum, using simulations from the OWLS project \cite[][]{2010schaye}. 
Not only is there an effect on the clustering of baryons by the material that is ejected out to large radii, the dark matter reacts to these changes by expanding, a phenomenon dubbed 'the back reaction' \cite[e.g.][]{duffy2010impact, 2011vanDaalen, velliscig2014impact}.
This suppression of the matter power spectrum has been confirmed and/or modelled by others since then \cite[e.g.][]{Vogelsberger2014, mead2015accurate, hellwing2016effect, Mummery2017, 2018Chisari, Schneider2019, Arico2021, pandey2023inferring, Salcido2023}.
%\cite{2020vanDaalen} expanded on this by considering a library of 92 cosmological simulations using different models and parameters, again finding AGN feedback to have a considerable impact on the matter power spectrum.
However, the mechanisms causing (AGN) feedback and the consequences on the surrounding material are as of yet inadequately understood in the context of the matter power spectrum to predict the power spectrum to near $1\%$ accuracy over a sufficient range of scales based solely on cosmology \cite[e.g.][]{2020vanDaalen}. 
%This is illustrated by the fact that the simulation parameters used to model it, which are further discussed in section \ref{sims}, are not fully optimised.%, making research in this area essential 
%[ander woord zoeken] \cite[][]{2010schaye, 2020vanDaalen}.

In order to make valid predictions of the power spectrum, we need to understand which parts of the matter distribution contribute significantly to the full power spectrum on any given scale.
One way to separate the power spectrum into components, is to distinguish between the contributions to the full power spectrum of auto-power produced by haloes only, auto-power generated by matter not in haloes and cross-power between halo and non-halo matter.
%One such separation of the power spectrum, consists of power produced by haloes only, as opposed to power generated by matter not in haloes or cross-power between halo and non-halo matter. 
The contribution of these components is influenced by the definition of halo boundary used in a scale-dependent way, as found by \cite{2015vanDaalen} for the halo auto-power spectrum in dark matter only (DMO) simulations.

Keeping a consistent definition for halo boundaries, e.g. a radius marking a spherical overdensity (SO) region, separate power spectra for haloes of certain masses can be considered, to explore their individual or combined effects to the full power spectrum. 
Therefore, \cite{2015vanDaalen} considered the contribution of power in haloes to the full power spectrum in DMO simulations for a varying minimum halo mass.
They found that group-sized haloes (with $M_{200,\mathrm{mean}}\gtrsim$\hM{13.75}) are the main contributors to matter clustering for $2\lesssim$\ksc{\lesssim}{10}, even though they contain only $\sim13\%$ of the total mass.
The deviation between the mass fraction and the contribution to the total power stems from the interplay between the rarity of massive haloes, lowering their contribution to the total power, and the deeper potential wells of massive haloes, which result in increased clustering of the matter, along with their already more clustered environments (i.e.\ their bias), raising their contribution to the total power \cite[see also e.g.][]{debackere2020impact}.
%Regardless, the contribution to the full power spectrum does appear to be converged with decreasing halo mass, at $2\lesssim$\ksc{\lesssim}{10}, as found by \cite{2015vanDaalen}. 
Even so, \citet{2015vanDaalen} found that a significant ($>1\%$) amount of power is still provided by unresolved haloes ($M_{200,\mathrm{mean}}\lesssim$\hM{9.5}) and, potentially, smooth dark matter.

The relevance of the large contribution of group-sized haloes to matter clustering was further confirmed when \cite{2020vanDaalen} examined a library of 92 power spectra for cosmological simulations using different galaxy formation models and parameters. They found that the mean baryon fraction of these ($M\sim$\hM{14}) haloes can be used to predict the power suppression due to galaxy formation to typically within $1\%$ for \ksc{<}{1}.

Considering the importance of group-sized haloes to clustering measurements and the studying of other physical processes \cite[e.g.][]{McCarthy2010,Oppenheimer_2021}, here we delve deeper into the role of these objects in shaping the matter power spectrum. To this end we utilize the cosmo-OWLS \cite[][]{lebrun2014, mccarthy2014thermal} and BAHAMAS \cite[BAryons and haloes of MAssive Systems,][]{McCarthy2017} simulations, which were made to study haloes of these masses, and try to quantify the impact of AGN feedback on the contribution of (group-sized) haloes to the matter power spectrum.
Furthermore, we explore the influence of AGN feedback on the suppression of power on smaller scales for different components of the power spectrum, creating a model, based on the key element of mass removal from clustered regions, that can predict this suppression to percent-level accuracy down to \ksc{\approx}{10} in a way that is independent of the galaxy formation recipes used.
By furthering our understanding of the contributions haloes of different masses have, and the effects of baryonic feedback processes, we hope this work can provide a stepping stone for future research to increase the precision of theoretical power spectrum predictions on non-linear scales.

This paper is structured as follows. In \S \ref{methods} we describe the simulations in more detail, and provide a formal definition of the power spectrum and other methods used to obtain the results. \S \ref{results} then describes the results, where the contributions of halo auto-power for an unmatched set of \radm haloes is considered in \S \ref{nonmatched}.
The halo auto-power for a matched set of haloes is similarly considered in \S \ref{matched}. Using this same matched catalogue, we also explore the clustering contributions of \radc SO regions instead (\S \ref{so_m}), consider the results when using a fixed SO radius between simulations (\S \ref{dm_rad}), and investigate the effects of varying the AGN feedback temperature (\S \ref{fb_temp}). The effects of cosmology on the results is briefly discussed in Appendix~\ref{cosm}. Other contributions to the power spectrum, i.e. cross terms, are considered in \S \ref{othercontributions}, based on which we derive a model for power suppression in \S \ref{modelmain}. Finally, \S \ref{summary} provides a summary of our results and a brief discussion of the model's potential.

\begin{table*}
    \begin{center}
\centering
\begin{tabular}{lllll} 
\hline
Name                       & Abbreviation &  $m_{\mathrm{dm}}[ h^{-1} \mathrm{M_\odot}]$ & $m_{\mathrm{bar, initial}}[ h^{-1} \mathrm{M_\odot}]$ & $\Delta T_{\mathrm{heat}}$  \\ 
\hline
C-OWLS\_DMONLY\_WMAP7               & C\_DMO\_W7      & $4.50\times10^9$                            & -                                                     & -                           \\
C-OWLS\_REF\_WMAP7                  & C\_REF\_W7      & $3.75\times10^9$                            & $7.54\times10^8$                                      & -                           \\
C-OWLS\_AGN\_WMAP7                  & C\_AGN\_W7      & $3.75\times10^9$                            & $7.54\times10^8$                                      & $10^{8.0} \mathrm{K}$       \\
BAHAMAS\_DMONLY\_2fluid\_nu0\_WMAP9 & B\_DMO\_W9      & $3.85\times10^9$                            & $7.66\times10^8$                                      & -                           \\
BAHAMAS\_nu0\_WMAP9                 & B\_AGN\_W9      & $3.85\times10^9$                            & $7.66\times10^8$                                      & $10^{7.8} \mathrm{K}$       \\
BAHAMAS\_Theat7.6\_nu0\_WMAP9       & B\_AGN7p6\_W9      & $3.85\times10^9$                            & $7.66\times10^8$                                      & $10^{7.6} \mathrm{K}$       \\
BAHAMAS\_Theat8.0\_nu0\_WMAP9       & B\_AGN8p0\_W9      & $3.85\times10^9$                            & $7.66\times10^8$                                      & $10^{8.0} \mathrm{K}$       \\
BAHAMAS\_DMONLY\_2fluid\_nu0\_Planck2013 & B\_DMO\_PL      & $4.44\times10^9$                            & $8.11\times10^8$                                      & -                           \\
BAHAMAS\_nu0\_Planck2013            & B\_AGN\_PL      & $4.44\times10^9$                            & $8.11\times10^8$                                      & $10^{7.8} \mathrm{K}$      
\end{tabular}
\caption{Relevant simulation parameters for simulations from BAHAMAS \citep[][]{McCarthy2017} and cosmo-OWLS \citep[][]{lebrun2014} that are used in this research. Simulation names (and their abbreviations) include the cosmologies used, along with particle types present (DMO only contains dark matter, REF and AGN are baryonic). All simulations have volumes $(400\,h^{-1}\,\mathrm{Mpc})^3$. The AGN simulations include thermal AGN feedback, with heating temperature $\Delta T_{\mathrm{heat}}$, REF only has SN feedback.}
\label{simparams}
    \end{center}
\end{table*}

\section{Methods}
\label{methods}
\subsection{Measurements and data}

\subsubsection{Simulations}
\label{sims}

The results in \S \ref{results} are based on a set of simulations from the cosmo-OWLS \citep[][]{lebrun2014} and BAHAMAS \citep[][]{McCarthy2017} simulation projects, which were designed to study group- and cluster-sized haloes. 
Both projects were run on a modified version of \gadget, with smoothed particle hydrodynamics (SPH), as described by \cite{springel2005cosmological}. 
The relevant simulation parameters are listed in Table~\ref{simparams} along with the abbreviations of the simulation names, which will be used in the rest of the paper to refer to the simulations. 

All simulations used in this study have a volume of $(400\, h^{-1}\,\mathrm{Mpc})^3$ and $1024^3$ particles ($\times 2$ for simulations including baryons, or 2-fluid DMO simulations). These are therefore generally indicated with the suffix L400N1024, but for brevity we will omit this suffix everywhere, except in Appendix~\ref{boxsize}, which studies boxsize effects on the results at a constant resolution.
The DMO simulations, as the name indicates, include only dark matter particles. In the REF simulation (cosmo-OWLS), gas and star particles and their relevant physics (e.g.\ star formation, metal enrichment, supernova feedback) have been added, and the AGN simulations add supermassive black hole particles along with their accretion and feedback to the mix, which are seeded into haloes with $>100$ dark matter particles.
The particle masses are listed in table \ref{simparams}.
The doubling of the number of particles, along with the mass difference between them in a DMO and a baryonic simulation, causes a variation in resolution effects between these simulations on small scales, which also affects the matter clustering.
Therefore, in the BAHAMAS project, the 2-fluid DMO simulations split every dark matter particle into two dark matter particles, with the dark matter-baryon mass ratio, and use a separate transfer function for each, to account for this effect and make a comparison on smaller scales between DMO and baryonic simulations more easily interpretable.

For the cosmo-OWLS simulations, the seven-year WMAP cosmological parameters are used \cite[WMAP7,][]{Komatsu_2011}, as listed in table \ref{cosmpar}. The BAHAMAS simulations use the newer, nine-year WMAP results \cite[WMAP9,][]{Hinshaw_2013}, also listed in table \ref{cosmpar}. To investigate whether our results show a dependence on cosmology, these simulations are compared to the simulations run with the 2013 Planck cosmological parameters \cite[][, named 'Planck' in table \ref{cosmpar}]{2014planck} in Figure~\ref{bah_mcm_cosm} of Appendix~\ref{cosm}.

\begin{table}
\centering
\begin{tabular}{lllllll} 
\hline
       & $\Omega_\mathrm{m}$ & $\Omega_\mathrm{b}$ & $\Omega_\Lambda$ & $\sigma_8$ & $n_s$  & $h$     \\ 
\hline
WMAP7  & 0.272      & 0.0455     & 0.728            & 0.81       & 0.967  & 0.704   \\
WMAP9  & 0.2793     & 0.0463      & 0.7207           & 0.821      & 0.972  & 0.700   \\
Planck & 0.3175     & 0.0490     & 0.6825           & 0.834      & 0.9624 & 0.6711 
\end{tabular}
\caption{Cosmological parameters from the seven-year WMAP release \citep[][]{Komatsu_2011}, the nine-year WMAP release \citep[][]{Hinshaw_2013} and the 2013 Planck release \citep[][]{2014planck} used in the simulations from table \ref{simparams}.}
\label{cosmpar}
\end{table}

Starting from initial conditions generated by V. Springel's software package \NGenIC\footnote{\url{http://www.mpa-garching.mpg.de/gadget/}}, which was updated to include second-order Lagrangian Perturbation Theory by S. Bird for BAHAMAS, and using a transfer function as prescribed by \citet{eisenstein1999power} (cosmo-OWLS) or the Boltzmann code \CAMB\footnote{\url{http://camb.info/}} \citep[][]{lewis2000efficient} (BAHAMAS), the particles in the simulations are evolved to $z=0$ by models, describing physical processes and 'sub-grid' models, describing unresolved, but relevant, physics \citep[for an extended description see][]{2010schaye, lebrun2014, McCarthy2017}.

Processes that are described by sub-grid models include baryonic feedback, i.e. supernova (SN) and AGN feedback. 
In their AGN simulations, cosmo-OWLS and BAHAMAS both use thermal AGN feedback, which entails that the feedback is 'accumulated' in the black hole until $1$ (cosmo-OWLS) or $20$ (BAHAMAS) surrounding particle(s) can be heated by $\Delta T_{\mathrm{heat}}$. 
This heating temperature is a model parameter, shown in table \ref{simparams}, that is varied between simulations, as its 'true' value is unknown. 
For the B\_AGN\_W9 simulation, the parameters governing AGN feedback have been fitted to multiple sets of observational data \cite[see][for details]{McCarthy2017}. 
The other BAHAMAS simulations, as well as the cosmo-OWLS AGN simulation, explore different feedback strengths, both weaker and stronger. The cosmo-OWLS 'reference' (REF) simulations include SN feedback but not AGN feedback, causing the well-known `overcooling' problem \citep[see][]{McCarthy2010, McCarthy2011overcool}.

\subsubsection{Power spectra}
\label{powerspectra}
Recall from \S \ref{intro} that the matter power spectrum can be seen as the amplitude squared of the Fourier transform of the matter overdensity field. Most of our results will be shown in terms of the power spectrum, defined as:
\begin{equation}
    P(k) = V \left< |\hat{\delta}_\textbf{k}|^2 \right>_\mathrm{k} ,
\end{equation}
where $k$ is the Fourier scale, $P$ is the power, $V$ the volume of the simulation (which is $(400 \,h^{-1}\mathrm{Mpc})^3$ everywhere) and $\hat{\delta}_\textbf{k}$ is the Fourier transform of the density contrast.
In Figure \ref{cos_m_nm_200}, an exception is made and $P(k)k^{1.5}$ is shown to take out the main $k$-dependence and make the comparison between lines easier. 
%Note that when ratios of $\Delta^2$ are depicted, the pre-factors drop out and essentially, a ratio of the density contrast amplitudes squared is shown.
The shot noise, generated by the `clustering' of particles with themselves, is subtracted from the power to account for the discreteness of the density field.

All auto-power spectra are calculated using a modified version of \powmes \citetext{as described in \citealp{2015vanDaalen}; see \citealp{colombi2009accurate} for the original version}, which takes as input a particle selection, along with a list of parameters and a path to the simulation data, and outputs the power at each scale.
%The binary files are created beforehand using the simulation data and a \python code that selects based on given criteria. 

The cross-power spectra, used in \S\ref{othercontributions} and \S\ref{modelmain}, are calculated using \nbodykit \cite[see][for a detailed description]{2018hand}. 
%The selection files made for \powmes are reused to create the data catalogs, containing particle positions and masses. \nbodykit subsequently calculates the cross-power between the two given datasets as a function of scale.

\subsubsection{Selections}
\label{selections}
For the results in \S \ref{results}, the \subfind output of the simulations is first used to divide the haloes based on mass, whereafter the bound and spherical overdensity particles are selected based on their radial distance to the centre of the halo, using \python.

The haloes are defined using cut-off radii where either \radm or \radc is chosen, corresponding to the radii wherein the average density is equal to $200$ times the mean density ($200\times\Omega_\mathrm{m}\rho_{\mathrm{crit}}$) or $500$ times the critical density ($500\times\rho_{\mathrm{crit}}$), respectively.
However, in order to enable comparison between figures, the halo mass we select on is always chosen as the mass within \radc (\M$ = 500\times\frac{4\pi}{3}\rho_{\mathrm{crit}}$\radc$\!\!^3$), regardless of the SO region used in particle selection. 
Furthermore, a cutoff of 100 particles is enforced to counter resolution effects. This mostly affects the lowest mass bin of figures where haloes are matched between simulations (see \S \ref{matching}) because not all haloes containing $>100$ particles have a $>100$ particle match.

The halo mass bins range from \M = \hM{11.25} to \M = \hM{14.75} with a width of 0.5 dex, to reduce mass evolution effects within the mass bin, while ensuring a large enough sample size. For the figures where haloes are matched between simulations (see \S \ref{matching}), the halo mass selection is based on the DMO simulation masses, contrary to the unmatched figures where the masses correspond to the simulation shown. 

\begin{figure}
\includegraphics[width=20pc]{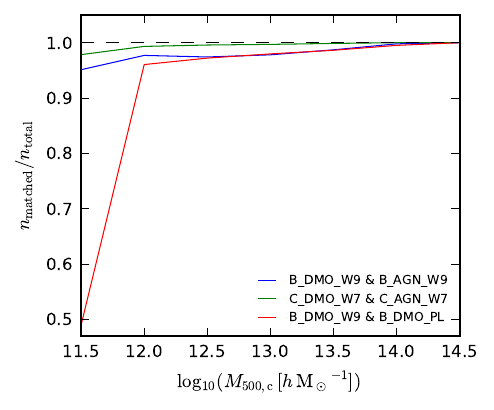}
\caption{The fraction of matched haloes out of the total number per halo mass bin (width 0.5 dex) for the match between B\_DMO\_W9 and B\_AGN\_W9 (blue), between C\_DMO\_W7 and C\_AGN\_W7 (green) and between B\_DMO\_PL and B\_DMO\_W9 (red).
With the exception of the lowest mass bin in the cosmology match (\M=\hM{11.5\pm0.25}), all match fractions are $>0.95$.
}
\label{match_frac}
\end{figure}

\subsection{Matching}
\label{matching}
For some figures in \S \ref{results} (e.g. Figure~\ref{cos_mcm_m_200}), the haloes are matched between simulations, which is made possible by the initial conditions of simulations in the same set being identical. For cosmo-OWLS, two sets of matches were made: haloes in C\_DMO\_W7 were matched with their equivalents in C\_AGN\_W7, and separately with their haloes in C\_REF\_W7, for each boxsize (L$100$, L$200$, L$400$), for halo masses \hM{11.25} and up.  
For BAHAMAS, the following sets of matches were made: B\_DMO\_W9--B\_AGN\_W9, B\_DMO\_W9--B\_AGN7p6\_W9, B\_DMO\_W9--B\_AGN8p0\_W9 and B\_DMO\_PL--B\_AGN\_PL, for halo masses $\geq$ \hM{11.25}.
A match between B\_DMO\_W9 and B\_DMO\_PL for halo masses $\geq$ \hM{11.25} was also made, to allow for a comparison between cosmologies (see Appendix~\ref{cosm}).

During the matching, the particle IDs of the $100$ most-bound dark matter particles of each halo within the mass range in the DMO simulation, are compared to \emph{all} the dark matter IDs of the haloes in the same or a neighbouring grid section in the other simulation. The halo where most of these particles are found is placed in a list, where a minimum of $50$ particles is required to be considered a match. The particles of the $100$ most-bound dark matter particles of \emph{these} haloes are then found in the DMO simulation in the same way. Furthermore, a halo position check ensures that the matched haloes are $<1\,\mathrm{Mpc}$ apart.
In the BAHAMAS matching, the computing time is sped up by also imposing a mass limit of a factor $5$ within the mass of the halo for haloes in neighbouring grid sections.
If a halo pair passes these tests it is considered a match and the index pair is saved.

The fraction of matched haloes as a function of halo mass is given in Figure~\ref{match_frac} for the match between B\_DMO\_W9 and B\_AGN\_W9 (blue line, representative for all BAHAMAS DMO-baryon pairs), the match between C\_DMO\_W7 and C\_AGN\_W7 (green), and the match between B\_DMO\_W9 and B\_DMO\_PL (red) used in \S \ref{matched}.
For \M$\geq$\hM{11.75}, the matched fraction is $>0.95$ for all pairs.
For the lowest mass bin (\M=\hM{11.5\pm0.25}) the matched fraction between different cosmologies is $\sim0.5$, calling for caution in the interpretation.

During the selection of particles for \powmes or \nbodykit (as described in \S \ref{selections}), the haloes are selected based on their DMO-simulation masses, and only matched haloes are considered. For the results in \S \ref{dm_rad}, the SO region radii from the corresponding haloes in the DMO simulation are taken as a selection criterion for the AGN simulation haloes.

\section{Results}
\label{results}
%\subsection{Auto-Power}
\label{autopower}

When considering the effects of galaxy formation on the theoretical matter power spectrum, it is common to consider the ratio of a power spectrum from a hydrodynamical simulation to that of a DMO simulation, so that small (but relevant) deviations from unity become more apparent \citep[see e.g. Figure 2 of][]{2020vanDaalen}. In general, the shape of this ratio is a dip downwards from unity for \ksc{\gtrsim}{0.1}, meaning the power in realistic simulations is suppressed on large and intermediate, non-linear scales, relative to a dark matter only universe, due to matter being removed from clustered regions by feedback processes. This dip flattens out to a suppression of $\sim 10-20\%$ depending on the model used (with more effective feedback causing a stronger suppression), followed by a sharp upturn for \ksc{\gtrsim}{10} into ratios far above unity on small scales. This upturn is caused by the cooling of the baryons into a central galaxy on even smaller scales, which increases the concentration of the inner dark matter halo.

To more precisely understand the physical processes behind the shape of the power spectrum and its suppression due to feedback, we need to dive deeper into the different contributions that make up the power spectrum. Therefore, this section focuses on the contributions of a range of halo masses to the full spectrum and the ratio between baryon and dark matter only simulations to gain more understanding about the role of feedback processes for different halo masses.

\subsection{Auto-power contributions for non-matched haloes}
\label{nonmatched}
%There are multiple ways of selecting haloes based on mass, as described in \S \ref{selections}.
In this section, the \M\, halo masses include baryons when present in the simulation. 
Furthermore, the simulations are not matched, resulting in skewed halo distributions in the mass bins between simulations, as AGN feedback lowers halo masses.
%Furthermore the simulations are not matched, which means that when comparing haloes in the same mass bins between simulations, these are not the same haloes because AGN feedback lowers the halo masses.
\S \ref{matched} describes the results that are obtained when the haloes are matched between simulations.
\\

\begin{figure}
%  \begin{center}
\includegraphics[width=1.03\columnwidth]{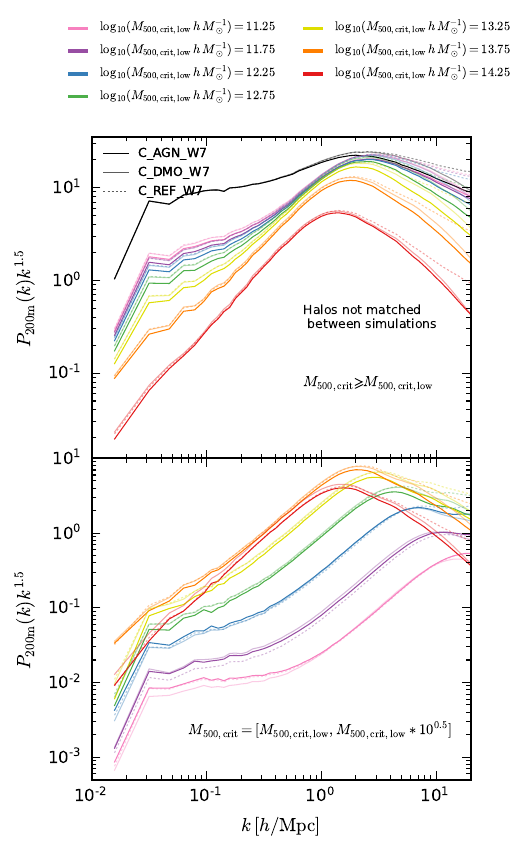}
\vspace{-0.7cm}
\caption{The power spectra of C\_AGN\_W7 (solid),  C\_DMO\_W7 (transparent) and C\_REF\_W7 (transparent and dashed) for various halo mass selections, at $z=0$.
The legend shows which \M\, selection is made: \M $\geq$ \hM{x-0.25} (top panel) or \hM{x-0.25} $\leq$ \M $\geq$ \hM{x+0.25} (bottom panel), with $x$ ranging from $11.5$ to $15$ (see top legend for colour scheme).
Only particles within the SUBFIND radius corresponding to a spherical overdensity of $200 \bar{\rho}$  ($r\!=\,$\radm) are taken into account.
The contribution of haloes with \M = \hM{14\pm0.25} to the full spectrum exceeds the contributions of haloes with other masses at \ksc{<}{3}. A comparison between simulations shows that AGN feedback lowers the fraction of mass residing in haloes of all masses, the SO radius, as well as the contribution to the full power spectrum on small scales (\ksc{>}{8}) of haloes with \M $\gtrsim$ \hM{11.75}.
%The dimensionless power spectra (left column) and mass contribution (fraction of full power spectrum of , right column) of C\_AGN\_W7\_L400N1024 (solid) and  C\_DMO\_W7\_L400N1024 (transparent) for haloes with $M_{500,crit} \geq 10^{x- 0.25}$ (top row) and $M_{500,crit}=10^{x \pm 0.25}$ (bottom row); $x$ ranging from 11 to 15 (see legend for colour scheme), at z=0.
}
\label{cos_m_nm_200}
%    \end{center}
\end{figure}

%\subsubsection{Mass contribution}
%\label{mc_nm}
In Figure~\ref{cos_m_nm_200} the power spectra of simulations C\_AGN\_W7 (solid),  C\_DMO\_W7 (transparent) and C\_REF\_W7 (transparent and dashed) are shown for a range of halo mass bins between \hM{11.25} and \hM{14.75} (with a width of 0.5 dex), at $z=0$.
The power spectra are multiplied by $k^{1.5}$ to take out the main $k$-dependence and to make a comparison between the lines easier. 

In the top panel, all particles within \radm of haloes with masses greater than that indicated in the legend are taken into account.
The more haloes are taken into account, the closer the power spectrum is to the full power spectrum (black line). Note that we do not expect to converge to this limit as more haloes are taken into account, since not all matter resides in (resolved, \radm) haloes.
For all mass bins, the power spectra have a similar shape: the power increases with $k$, reaches a maximum at the scale at which the (larger) haloes are entered and then decreases (keeping in mind that we multiply $P(k)$ with $k^{1.5}$ here). The scale at which the haloes are entered becomes larger with increasing halo mass, effectively moving the peak in the figure towards smaller $k$.

A comparison of simulations reveals a lower power of C\_AGN\_W7 on large scales, compared to both C\_DMO\_W7 and C\_REF\_W7, due to AGN feedback lowering the halo mass. The haloes that are in a certain C\_AGN\_W7 mass bin, are in in a higher mass bin in the C\_DMO\_W7 or C\_REF\_W7 simulations. As low-mass haloes are more numerous than high-mass haloes, more haloes will be included in the C\_DMO\_W7 and C\_REF\_W7 mass bins, increasing the power.
On smaller scales, C\_REF\_W7 power increases more strongly due to over-cooling while C\_AGN\_W7 dips under C\_DMO\_W7 as the effective feedback heats the gas and pushes baryons to larger scales, the dark matter reacting gravitationally.

In the bottom panel of Figure~\ref{cos_m_nm_200}, only particles within \radm of haloes within the indicated mass bin (width of 0.5 dex) are taken into account.
The shapes are similar to the top panel, although a comparison between masses reveals that the power on large and intermediate scales increases with halo mass for haloes with \M $\lesssim$ \hM{14.25}. The power of the largest haloes (\M=\hM{14.5\pm0.25}, red lines) is lower or equal to the second largest mass bin (\M=\hM{14\pm0.25}, orange lines), as was found previously by \citet{2015vanDaalen}. Although cosmic variance plays a minor role in this (as we checked using three B\_AGN\_W9 simulations with varying initial conditions, not shown), it is mostly due to cosmic deficiency. In other words, the \hM{14.5\pm0.25} haloes are so rare that their higher mass and bias no longer wins out, lowering their contribution to the full power spectrum relative to slightly less massive haloes.

For low-mass haloes (\M $\lesssim$ \hM{13}), the peaks of the auto-power spectra from the baryonic simulations lie above those from C\_DMO\_W7, whereas those for higher mass haloes lie below. The transition happens around the mass when AGN feedback becomes important, supporting the statement that AGN feedback is the cause.
Comparing the bottom to the top panel, we see that C\_AGN\_W7 lies below C\_DMO\_W7 when all haloes are taken into account, showing again that what happens to clustering in the groups and clusters dominates the effect on the overall power.

%\begin{figure}[h!]
%\includegraphics[width=35pc]{Figures/nonmatched/r500c/psdl_WMAP7_L400N1024_032_masses.pdf}
%\caption{
%The dimensionless power spectra (left column) and mass contribution, fraction of full power spectrum of corresponding simulation, (right column) of $C\_AGN\_W7\_L400N1024$ (solid),  $C\_DMO\_W7\_L400N1024$ (transparent) and $C\_REF\_W7\_L400N1024$ (transparent and dashed), at $z=0$.
%The legend shows which \M selection is made: $M_{500,crit} \geq 10^{x- 0.25}$ (top row) and $M_{500,crit}=10^{x \pm 0.25}$ (bottom row); $x$ ranging from 11 to 15 (see legend for colour scheme).
%Figure is as figure \ref{cos_m_nm_200} with the exception that
%only particles within the SUBFIND radius corresponding to a spherical overdensity of $500 \rho_{crit}$  ($r=r_{500,crit}$) are taken into account.
%The contribution of haloes with $M_{500,crit}=10^{14 \pm 0.25}\, \mathrm{M_\odot/h}$ to the full spectrum exceeds the contributions of haloes with other masses at $k<7\,h\,Mpc^{-1}$. A comparison between simulations shows that in the inner regions of haloes with $M_{500,crit}>10^{12.75}\, \mathrm{M_\odot/h}$, either mass is lost or the SO radius is reduced in the presence of AGN feedback. 
%}
%\label{cos_m_nm_500}
%\end{figure}

\begin{figure}
%  \begin{center}
\includegraphics[width=1.03\columnwidth]{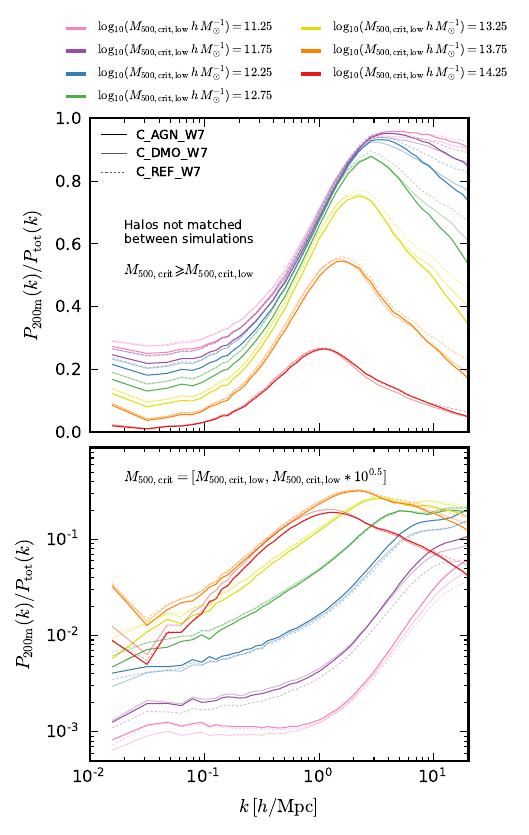}
\vspace{-0.7cm}
\caption{The halo clustering contribution as a function of mass, relative to the full matter power spectrum, for C\_AGN\_W7 (solid),  C\_DMO\_W7 (transparent) and C\_REF\_W7 (transparent and dashed), at $z=0$.
The legend and particle selections are the same as in Figure~\ref{cos_m_nm_200}.
%Only particles within the SUBFIND radius corresponding to a spherical overdensity of $200 \bar{\rho}$  ($r=r_{200,mean}$) are taken into account.
The contribution of haloes with \M = \hM{14\pm0.25} to the full spectrum exceeds the contributions of haloes with other masses for \ksc{<}{3}. 
AGN feedback lowers the fraction of mass residing in massive haloes (\M$\gtrsim$\hM{13}), as well as the contribution to the full power spectrum on small scales \ksc{\gtrsim}{8} of haloes with \M $\gtrsim$ \hM{11.75}, though resolution effects start to play a significant role at this scale.
%The dimensionless power spectra (left column) and mass contribution (fraction of full power spectrum of , right column) of C\_AGN\_W7\_L400N1024 (solid) and  C\_DMO\_W7\_L400N1024 (transparent) for haloes with $M_{500,crit} \geq 10^{x- 0.25}$ (top row) and $M_{500,crit}=10^{x \pm 0.25}$ (bottom row); $x$ ranging from 11 to 15 (see legend for colour scheme), at z=0.
}
\label{cos_mc_nm_200}
%    \end{center}
\end{figure}

To better compare the relative contributions of haloes between the different simulations, it is useful to consider ratios of power spectra instead. In Figure~\ref{cos_mc_nm_200}, each of the lines portrayed in Figure~\ref{cos_m_nm_200} is divided by the full power spectrum of the accompanying simulation to illustrate the contribution to the full spectrum for each halo mass bin. The masses and line styles (indicating simulations) are as in Figure~\ref{cos_m_nm_200}. Here we still take \radm as the halo boundary.

On large scales, the contribution of haloes to the power flattens off to a fraction of $\approx$ 0.3 (depending on simulation) when all haloes are taken into account (pink lines). In this regime, the 2-halo term dominates: the overdensities are determined by the positions and masses/sizes of the haloes, the clustering in the interior of haloes has a negligible contribution.
As we move to higher $k$ (smaller scales), there is a massive rise in contribution around the scales that haloes are entered, which of course differs per mass bin. 
Along this rise, the 1-halo term starts to dominate over the 2-halo term, as most of the power comes from clustering in the interior of the haloes. 
Once the halo is entered, a maximum power contribution to the total is reached, after which it declines again for smaller scales. 
Note that for \ksc{>}{8} resolution effects start to influence the results noticeably, lowering the power.
When all haloes are taken into account (pink line, top panel of Figure~\ref{cos_mc_nm_200}), the contribution to the full power spectrum reaches $>0.9$ at \ksc{\approx}{3}. 
The remaining percentage of power is provided by a combination of the cross-power between non-halo and halo particles and the auto-power in non-halo particles (as seen in \S \ref{othercontributions}).

The bottom panel of Figure~\ref{cos_mc_nm_200} shows the individual contributions of each halo mass bin to the full power spectra of C\_AGN\_W7, C\_REF\_W7 and C\_DMO\_W7. This panel clearly shows that the \hM{14 \pm 0.25} haloes contribute most to the full power spectrum for \ksc{<}{3}, and lower mass bins take over on smaller scales. Note that resolution effects play some role in this, which generally lower the power on small scales (\ksc{\gtrsim}{8}).

On large scales, the fraction of the total mass selected, when selecting the haloes in a certain mass bin, determines the fraction of power to the full power spectrum. 
As the selection is purely based on halo mass, one might expect all three simulations to lie on top of each other. 
However, because these are not the same haloes, this is not the case; the discrepancies become larger, as more haloes are taken into account (top panel, \ksc{\sim}{0.02}). While C\_REF\_W7 and C\_DMO\_W7 are in quite good agreement, the total halo contribution on large scales is significantly lower in C\_AGN\_W7, as more mass is removed from haloes.
Looking at the peak in contribution (when entering the halo) in the top panel of Figure~\ref{cos_mc_nm_200} for masses around \hM{13.75} (yellow and orange lines), there is a clear horizontal displacement between the power contribution of haloes to the full power spectrum in the C\_AGN\_W7 and C\_DMO\_W7. This is counter-intuitive, since AGN feedback lowers the halo mass, effectively placing the more extended, denser haloes in a lower mass bin, from which one would expect the peak in C\_AGN\_W7 to move to larger scales (smaller $k$). Looking at the bottom panel, however, we see that the shift does not occur in individual halo mass bins, but is a cumulative effect of feedback reducing the halo mass by different amounts on different mass scales.

We also see that the peak of these haloes in C\_REF\_W7 is considerably higher than both other simulations, likely due to extra cross-power between halo particles and the large number of stars that have formed due to overcooling.

When comparing the C\_AGN\_W7, C\_REF\_W7 and C\_DMO\_W7 lines in Figure~\ref{cos_mc_nm_200}, drawing conclusions is not straightforward because the haloes are not matched between simulations. Baryon cooling strongly peaks the density distribution towards the center of the haloes, changing SO radii and thus the scale at which the halo is entered. Additionally, the removal of mass due to feedback also changes the profiles and therefore the 1-halo contribution, while also lowering the overall power of the haloes and shifting haloes between mass bins. To be able to better distinguish these different effects, in the remainder of \S\ref{results} we will consider the relative contributions for matched catalogues of haloes.
%On small scales (\ksc{\sim}{6}), C\_AGN\_W7 dips below C\_DMO\_W7 for halo masses $\gtrsim$ \hM{12.75} (bottom  panel, green, yellow, orange lines), indicating that the feedback pushes material out of the center, as confirmed in \S \ref{matched}. C\_REF\_W7 rises above C\_DMO\_W7 and C\_AGN\_W7 due to the over-cooling of baryons on small scales, which leads to star formation and clumping of dark matter.
%\\

\subsection{Auto-power contributions for matched haloes}
\label{matched}
%As shown by \S \ref{nonmatched}), in order to make interpretable comparisons between the baryonic and DMO simulations, their haloes need to be matched. Fortunately, both cosmo-OWLS and BAHAMAS have been set up so this is a possibility: the simulations within the projects have the same initial conditions. Each baryon simulation is matched to the corresponding DMO simulation (keeping cosmology etc. the same between the matches) according to \S \ref{matching}. The fraction of matched haloes can be viewed in Figure~\ref{match_frac}. The halo mass selection is based on the DMO-simulation masses. 
%\\

\begin{figure*}
    \begin{center}
\includegraphics[width=1.0\textwidth]{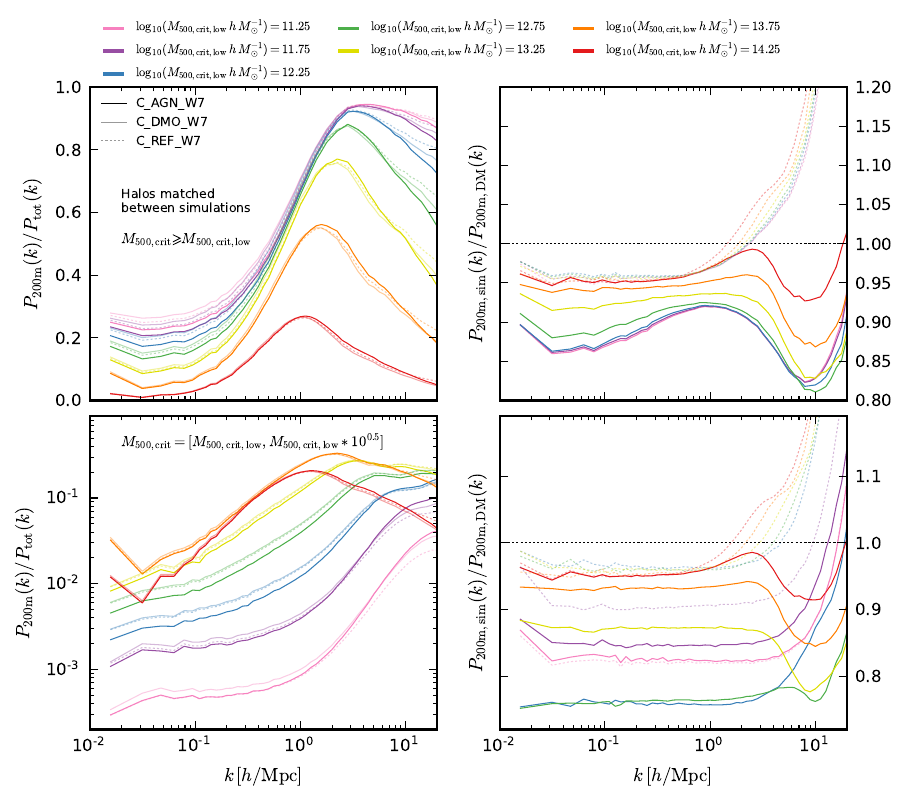}
\vspace{-0.7cm}
\caption{The left side of the figure is as in Figure~\ref{cos_mc_nm_200}, except that the haloes are matched between C\_DMO\_W7 and the baryonic simulations. Only matched pairs are shown. The right side of the figure shows the power spectra of the C\_AGN\_W7 (solid) and C\_REF\_W7 (transparent and dashed) simulations divided by the power spectra of C\_DMO\_W7, for each halo mass bin.
\M\, masses are measured in C\_DMO\_W7.
AGN feedback lowers the masses of all haloes. On small scales, AGN feedback causes a dip in the C\_AGN\_W7/C\_DMO\_W7 power ratios of massive (\M$\geq$\hM{12.75}) haloes, whereas the lower-mass haloes shown a sharp rise, like all C\_REF\_W7/C\_DMO\_W7 power ratios.}
\label{cos_mcm_m_200}
    \end{center}
\end{figure*}

%\subsubsection{Mass contribution}
%\label{mc_m}
The left side of Figure~\ref{cos_mcm_m_200} shows the power contribution to the full power spectrum for the same mass selections of haloes as in Figure~\ref{cos_mc_nm_200}, but for a matched set of haloes. 
The right side shows the ratios between the power spectra of the baryon and DMO simulations (C\_AGN\_W7/C\_DMO\_W7, solid continuous and C\_REF\_W7/C\_DMO\_W7, dashed transparent line style) for each mass bin, to illustrate the effect of baryons on the power in haloes for a range of halo masses.

On large scales, small $k$, the power contribution to the full power spectrum reflects the mass fraction of the selected haloes.
Therefore, the top-left panel shows that the masses of the haloes are not the same in the two simulations. 
When taking only the largest haloes into account (\M$\geq$\hM{14.25}, red lines), the masses in the baryon simulations are closest to those in C\_DMO\_W7, as seen in the top-right panel at \ksc{\approx}{0.02}, where the power ratio C\_AGN\_W7/C\_DMO\_W7 $\approx0.96$. 
These haloes have such a deep potential well that AGN feedback is damped and does not have a substantial effect. 
As more lower-mass haloes are added to the sample, the power ratio between simulations lowers until it saturates at $\approx0.88$ for \M $\geq$ \hM{12.25}, where the bottom-right panel shows that haloes with masses lower than \hM{12.75} either have a similar or a higher C\_AGN\_W7/C\_DMO\_W7 ratio and the bottom-left panel shows that these mass bins have a much lower contribution to the full power spectrum than the more massive haloes. 
Consequentially, the C\_AGN\_W7/C\_DMO\_W7 ratios of the mass bins that take lower mass haloes into account as well as higher mass ones, are determined by the C\_AGN\_W7/C\_DMO\_W7 ratios of the higher mass haloes up to $\sim$\hM{14} and thus lowered by the AGN feedback processes these are subject to. 

A possible reason why the simulation ratios of low mass haloes, e.g.\ \hM{12\pm0.25} (solid purple line, Figure~\ref{cos_mcm_m_200}), in the bottom-right panel are lower for AGN than REF, could be that these haloes are still indirectly effected by AGN feedback. Lower-mass haloes tend to cluster around high-mass ones, forming large groups. 
The powerful shock waves caused by AGN in massive haloes can carry out to far beyond these haloes, affecting all the other haloes in the vicinity. 
This might cause the haloes to lose all their gas due to a shock wave, or in milder cases, the expansion of the central halo weakens its gravitational pull making surrounding objects less strongly bound/expand. 
A more likely explanation is that the formation and evolution of the haloes were influenced by the AGN at an earlier stage. The gas around galaxies with AGN is much hotter, hindering the accretion and growth of smaller haloes and galaxies. However, this has not been verified in this research.
Either way, the effects of AGN feedback are felt by lower-mass galaxies, as can be seen when comparing the solid (C\_AGN\_W7/C\_DMO\_W7) and transparent dashed (C\_REF\_W7/C\_DMO\_W7) lines in the bottom right panel of Figure~\ref{cos_mcm_m_200}. Note that the C\_REF\_W7/C\_DMO\_W7 ratio is also $<1$, due to gas pressure which impedes collapse.
For low-mass haloes, \M$<$\hM{11.75}, there are few particles making up the halo in these simulations, allowing small discrepancies to have a large influence on the power. We also note that \citet{2015vanDaalen} found that such haloes are underrepresented at this resolution.

At $k$ of a few, the power spectrum is almost completely determined by the clustering of matter inside haloes, as can be seen by the peak in the top-left panel of Figure~\ref{cos_mcm_m_200}. Although the figure exhibits the same general shapes as the version without matching (Figure~\ref{cos_mc_nm_200}), considering matched rather than mass-selected catalogues has significant effects. 
The different simulations meet at a point, just before the peak, when the halo is entered; this is expected as they are the same haloes, although this would cease to happen if larger amounts of mass were removed. 
Right after this point, the contribution of haloes to the power in C\_AGN\_W7 rises above that of C\_DMO\_W7 and C\_REF\_W7. 
This however does not mean that halo matter in the C\_AGN\_W7 simulation clusters most strongly on this scale, compared to the other simulations: in fact, the \emph{opposite} is true. This effect comes from the differences in the full power spectra (through which we are dividing on the left-hand side of the figure), as is verified in the top-right panel of Figure~\ref{cos_mcm_m_200}. Here we see that the power ratio between C\_AGN\_W7 and C\_DMO\_W7 (solid lines) is below unity for every mass bin. It is therefore likely the cross-power between matter around haloes and haloes themselves being decreased by the presence of AGN which causes the rise of the peak seen for C\_AGN\_W7 in the top-left panel, due to the AGN removing mass from the haloes and depositing it on larger scales, where it clusters less strongly. %important point that was added

%On these scales, the power in C\_AGN\_W7 is lower than C\_DMO\_W7 because AGN feedback heats the gas, preventing clustering, and expels material outside the galaxy, decreasing the density and power. 
%This feedback expels material from the center to outside the galaxy, causing a back reaction from the dark matter, that also expands, lowering the density and power. 
%The C\_REF\_W7/C\_DMO\_W7 ratio starts to rise above $1$ at this scale because the dark matter haloes contract, reacting to the heavily clustered galaxy, and there is no AGN feedback and the supernova feedback is too weak to counter the contraction. 
%This confirms that the $<1$ C\_AGN\_W7/C\_DMO\_W7 ratio is not caused by other baryon effects. 
%Recall that the lack of sufficient feedback in C\_REF\_W7 results in overcooling, star formation and clumping, giving an enormous rise in power compared to C\_DMO\_W7 (transparent and dashed lines in Figure~\ref{cos_mcm_m_200}) at \ksc{<}{1}.

Still keeping our focus on the top-right panel, we see that, as expected, the relative auto-power contributions of haloes in different mass ranges compared to C\_DMO\_W7 is lower for C\_AGN\_W7 than for C\_REF\_W7, and that this difference is driven on large scales by the mass difference of $\sim$\hM{12.5}-\hM{14} haloes between these simulations. 
On small scales (at \ksc{\geq}{3}), where the change in auto-power is a reflection of the change in halo profile, the top-right panel shows a dip for all halo masses in the C\_AGN\_W7/C\_DMO\_W7 ratio. 
The C\_REF\_W7/C\_DMO\_W7 ratio rises above $1$ at this scale because the dark matter haloes contract, reacting to the heavily clustered galaxy, and there is no AGN feedback to counter the contraction (the SN feedback being too weak to counter the contraction on the most relevant mass scales). 
%This confirms that the $<1$ C\_AGN\_W7/C\_DMO\_W7 ratio is not caused by other baryon effects. %, allowing for the conclusion that AGN feedback is responsible for the dip in power ratios. 

At $3<k<10\,h\mathrm{\,Mpc^{-1}}$, the C\_AGN\_W7/C\_DMO\_W7 power ratio for \M $\geq$ \hM{12.75} haloes drops by $\approx 0.1$, because of AGN feedback. At \ksc{\geq}{10}, this ratio starts to rise again for all mass bins in the top-right panel, until, at \ksc{=}{20} each reaches more or less the value it had at \ksc{\approx}{3}. This rise can be attributed to the reaction of dark matter to the behaviour of baryons at small (galaxy) scales (\ksc{\sim}{100}), which cluster and increase the power, causing a contraction in the dark matter at larger scales (\ksc{\sim}{20}), meaning the AGN effectively change clustering the most at intermediate scales inside haloes, relative to DMO.
Despite their low contribution to the full power spectrum (bottom left panel), the lower mass haloes have a part to play in the rise at \ksc{\gtrsim}{10}, shown by the steeper slope of the C\_AGN\_W7/C\_DMO\_W7 power ratios at these scales in the top right panel for the lines that take the lower-mass haloes into account. 
The behaviour of the lower mass galaxies (\M$<$ \hM{12.75}) can be explained by their lack of, but vicinity to, galaxies with AGN feedback: they are not affected internally, since they do not host AGN themselves, but their masses are lower than they would have been in a universe without AGN. The solid C\_AGN\_W7/C\_DMO\_W7 ratios in the bottom right panel follow the same general shapes and have similar slopes as the transparent dashed ones (C\_REF\_W7/C\_DMO\_W7), albeit shifted horizontally, indicating similar physical effects. 
%Likely, this is because the AGN in the vicinity will affect (lower) the mass of the entire low mass halo, unrelated to scale, and the density profile of the haloes, which determines the shape and slope of the power ratio, remains unaffected. 
%However, this is speculation and to establish it, more research is needed.

We note that resolution effects, coming from the limited number of particles making up a halo, which prohibit measuring the clustering on very small scales and start to play a role at \ksc{\gtrsim}{8}, lowering the power relative to a higher-resolution simulation. This affects the DMO haloes in cosmo-OWLS most strongly, as these were run with half as many particles as the baryonic simulations, and therefore may affect the ratios shown here. 
%The resolution effects will be smaller for C\_AGN\_W7 than for C\_DMO\_W7 because there are more particles in C\_AGN\_W7 (gas, stars and black holes) that can contribute to the power, in the cosmo-OWLS simulations. 
However, the BAHAMAS simulation project uses 2-fluid simulations for DMO, so this effect is compensated (see \S \ref{sims}). As we will see in \S\ref{fb_temp}, Figure~\ref{bah_mcm_m_200} shows that the ratios still exhibit the dip and rise for the BAHAMAS simulations, albeit somewhat differently, confirming that resolution effects are not the leading cause of the small-scale effects presented here.

%Comparing the bottom left to the bottom right panel, it becomes clear that the differences in full power spectra cause the power contribution of haloes in C\_REF\_W7 to the full power spectrum to be lower than that of haloes in C\_DMO\_W7 on small scales, as their ratio (C\_REF\_W7/C\_DMO\_W7) far exceeds $1$. 
%The opposite happens with C\_AGN\_W7: at \ksc{\geq}{4}, the contribution to the total of lower mass haloes, e.g. \M = \hM{12\pm0.25} (purple solid line), exceeds the DMO contribution to the total, whereas their ratio (C\_AGN\_W7/C\_DMO\_W7) is still below $1$, because the full AGN power spectrum is lower than DMO. \\

\begin{figure*}
    \begin{center}
\includegraphics[width=1.0\textwidth]{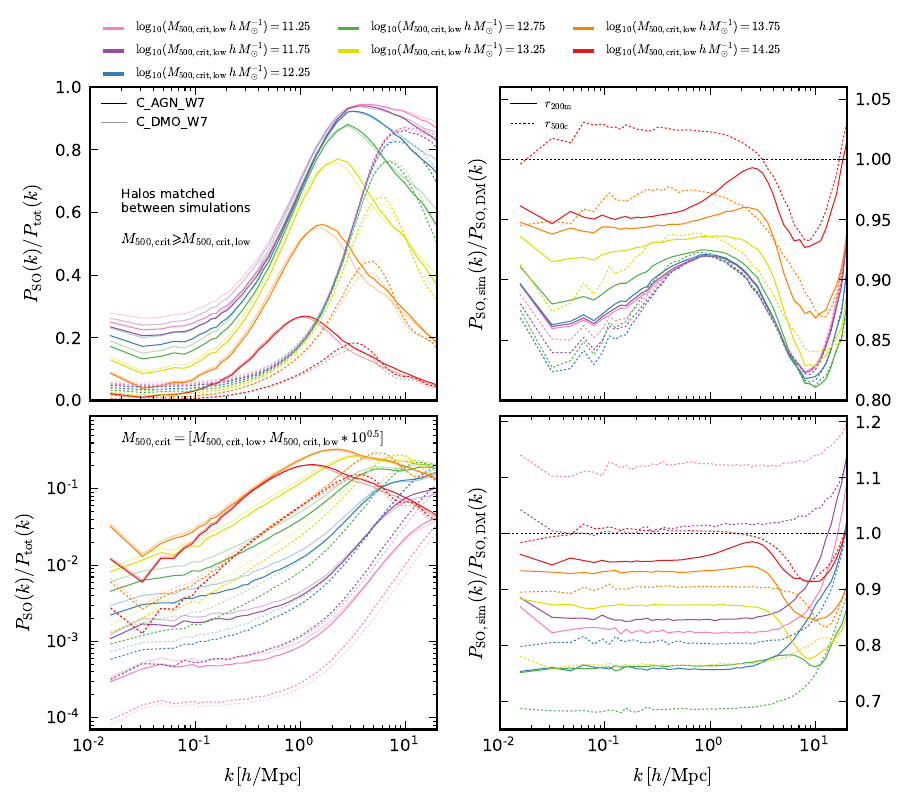}
\vspace{-0.7cm}
\caption{Absolute and relative halo power contributions for C\_AGN\_W7 and C\_DMO\_W7 as in figure~\ref{cos_mcm_m_200}, but dashed lines now show the results when only particles within \radc are included.
%The dimensionless power spectra (left column) and mass contribution, fraction of full power spectrum of corresponding simulation, (right column) of $C\_AGN\_W7\_L400N1024$ (solid) and  $C\_DMO\_W7\_L400N1024$ (transparent), at $z=0$.
%haloes are matched between DMO and the other simulations. Only matched pairs are shown.
%The legend shows which \M selection is made: $M_{500,crit} \geq 10^{x- 0.25}$ (top row) and $M_{500,crit}=10^{x \pm 0.25}$ (bottom row); $x$ ranging from 11 to 15 (see legend for colour scheme). Masses are measured in C\_DMO\_W7.
%The particles within the SUBFIND radius corresponding to a spherical overdensity of $200 \bar{\rho}$  ($r=r_{200,mean}$) (continuous linestyle) or $500 \rho_{crit}$  ($r=r_{500,crit}$) (dashed linestyle) are taken into account.
The peak in halo power contribution is moved to higher $k$ for particles within \radc\!, relative to the results for \radm\!. Although much less mass is included in the smaller \radc regions, the power on small scales only drops slightly, as they are highly biased.
As the bottom-left panel shows, for both SO regions $\sim$\hM{14} haloes provide the largest contribution to the total power up to $k\sim$ a few.
From the bottom-right panel, we see that for group-sized haloes the effects of AGN feedback are more extensive within \radc\!, compared to within \radm\!.
}
\label{cos_mcm_m_so}
    \end{center}
\end{figure*}

%\subsubsection{SO regions}
\subsection{Varying the SO regions}
\label{so_m}
We now consider how the halo contributions depend on the choice of halo boundary. Figure~\ref{cos_mcm_m_so} is similar to Figure~\ref{cos_mcm_m_200} for C\_AGN\_W7 (solid) and C\_DMO\_W7 (transparent), but dashed lines now show the results for a \radc cut-off radius (dashed) compared to \radm (continuous). The mass selections are as in Figure~\ref{cos_mcm_m_200} and the haloes are matched between the simulations.
%The right side of Figure~\ref{cos_mcm_m_so} shows the ratio between the C\_AGN\_W7 and C\_DMO\_W7 power spectra for the halo mass bins (as in Figure~\ref{cos_mcm_m_200}). 

The most drastic effect of the change in the SO radius from \radm to \radc on the relative halo contribution to the total power spectrum (top-left panel of Figure~\ref{cos_mcm_m_so}) is a horizontal shift of the peaks towards larger $k$. The \radc spherical overdensity region has a smaller radius than \radm and is often used to explore halo properties through X-ray observations. Direct or indirect effects from baryons are often more apparent when considering this region. Since the peak indicates the scale at which the halo is entered, the shift seen here is expected. 

Furthermore, on large scales the power contribution to the full power spectrum is lower for \radc\!, as less of the mass is taken into account in these calculations. 
% Similarly, on smaller scales, the contribution to the full power spectrum, when taking all haloes into account (top panel, pink dashed line), is lower for \radc than for \radm (pink continuous line), as we are not taking all the mass within the haloes into account.
Still, taking C\_DMO\_W7 as an example, the particles within \radc of haloes contribute a maximum fraction of $\approx0.89$ of the full power through their auto-power alone, while the particles within \radm reach $\approx0.96$ of the full power. In contrast, on large scales, \ksc{\sim}{0.02}, these SO regions account $\approx0.06$ of the total for \radc and for $\approx0.29$ of the total for \radm\!. Since the difference in power between SO regions on the largest scales is just a reflection of their relative total mass fractions, this indicates that the \radc regions contain slightly less than half the mass of the \radm regions. We thus see that almost all power on small scales comes from clustering within the \radc regions of haloes, even though these regions include much less mass. 

The bottom-left panel of Figure~\ref{cos_mcm_m_so} shows that for both SO radii, the \hM{14 \pm0.25} haloes provide the largest contribution to the full power spectrum for \ksc{\lesssim}{3}, although the dominance of these haloes continues up to \ksc{\sim}{7} for particles within \radc\!.
% %Apart from these discrepancies, the \radc results generally agree with the \radm results in shape.

On scales that are inside the halo ($k>k_{\mathrm{peak}}$), the \radc halo power spectra exhibit a slight oscillation that is caused by the sharp boundary of \radc\!, particularly in the bottom panel. This is roughly equivalent to the oscillations seen in the Fourier transform of a top-hat filter, and only more apparent for \radc than for \radm due to the higher density at the cut-off radius, and is therefore not a physical effect.

%As in Figure~\ref{cos_mcm_nm_so} (where the haloes were not matched), on large scales, the two SO regions in the top left panel of Figure~\ref{cos_mcm_m_so}, reflect the mass fraction squared present in the selection of particles, where the mass fraction within \radc is $\frac{2}{5}\Omega_\mathrm{m}$ of the mass fraction within \radm. 
Comparing C\_AGN\_W7 and C\_DMO\_W7 in the top-left panel of Figure~\ref{cos_mcm_m_so}, we see that the differences in halo contributions are larger around the peak for \radc than for \radm\!. Specifically, when baryons and AGN are included the peak auto-power contribution of haloes to the matter power spectrum is increased -- however, the top-right panel shows that this is again due to the total matter power spectrum decreasing in AGN relative to DMO, rather than the halo power itself. Like in \S\ref{matched}, we thus conclude that the increase in the relative peak power contributions are due to a decrease in cross-power of haloes with the matter around it. We consider the role of the cross-power further in \S\ref{othercontributions} and \S\ref{modelmain}.

Comparing the \radm and \radc results in the top-right panel further, we also see that the ``bump'' seen around \ksc{\approx}{3} for \radm for the two most massive halo mass bins (red and orange lines) is not seen for \radc\!. This bump is due to the mass moved to the outskirts of the halo by AGN feedback, and the fact that it is not visible for \radc means that a significant fraction of the mass is being deposited between \radc and \radm in the most massive haloes. The bottom-left panel shows that this is mostly due to the contribution of the very most massive haloes probed here (red line).
%Regarding the particles within \radc (dashed), the top right panel of Figure~\ref{cos_mcm_m_so} indicates that the C\_AGN\_W7/C\_DMO\_W7 ratios of the particles within \radc match the general trends of those within \radm. 
At the same time, from the same panel we see that the $\sim$\hM{14.5} haloes show a higher C\_AGN\_W7/C\_DMO\_W7 ratio for particles within \radc than particles within \radm on large scales, around unity for \ksc{\lesssim}{3}. This indicates that the mass added to \radc by the cooling of baryons is balanced by the mass ejected to $r>$\radc by AGN feedback for these haloes -- keeping in mind that the sizes of these regions may have changed between these simulations as well (see \S\ref{dm_rad}). We will refer to the region between \radc and \radm as the halo annulus, and consider it again in \S\ref{modelannulus}.
%For these masses the C\_AGN\_W7/C\_DMO\_W7 ratios also drop at higher $k$ for \radc than \radm, because the haloes are entered at higher $k$. 
%When lower-mass haloes are taken into account, the C\_AGN\_W7/C\_DMO\_W7 ratio for \radc is lower on large scales, indicating more mass loss, but joins up with the \radm selection ratios at \ksc{\sim}{1}. 

%This approximate agreement of the C\_AGN\_W7/C\_DMO\_W7 ratios in the two SO regions is not present in the lower right panel of Figure~\ref{cos_mcm_m_so}, where the C\_AGN\_W7/C\_DMO\_W7 ratios for the separate halo mass bins are shown. 
Looking at the other halo mass bins shown in the bottom-right panel of Figure~\ref{cos_mcm_m_so}, we see that for \M$\geq$\hM{12.75} (green, yellow and orange lines), the C\_AGN\_W7/C\_DMO\_W7 ratios are much lower for \radc than for \radm\!. This is expected, as mass is removed from the central regions by AGN feedback and deposited on larger scales, thereby affecting the mass within \radc more than that within \radm\!.
%This effect is reversed for haloes with \M$<$\hM{12.75}. 
%Considering the turnover mass is also the mass at which the haloes are large enough to host an AGN, one can assume that a significant portion of matter ejected from within \radc for \M$\geq$\hM{12.75} haloes is deposited between \radc and \radm; the rest is likely ejected beyond this. 
%The C\_AGN\_W7/C\_DMO\_W7 ratio on large scales for these masses is smaller for particles within \radc than for particles within \radm, leading to the theory that most of the mass will be deposited within \radm.
Contrarily, for \hM{12.5\pm0.25} haloes (blue), the C\_AGN\_W7/C\_DMO\_W7 ratio for \radc lies above that for \radm\!, but is still below unity indicating that the entire halo is less massive due to impeded growth or stripping of gas, as mentioned previously, and most of the mass is taken from/not added to the outer region of the halo.
For even less massive haloes (pink and purple lines), the C\_AGN\_W7/C\_DMO\_W7 ratio even exceeds unity for \radc\!, indicating that the cooling of baryons dominates over their ejection.
%most of the mass that is ejected, is probably deposited outside \radm, due to its shallower potential well.

Note that the SO radii change between the simulations as the halo mass changes, limiting the conclusions that can be drawn from this figure concerning where mass is deposited.
In \S \ref{dm_rad}, the radius is kept constant between simulations, to investigate this in more depth.

Finally, we note that the halo auto-power ratios between C\_AGN\_W7 and C\_DMO\_W7 for all mass bins as shown in the bottom-right panel of Figure~\ref{cos_mcm_m_so} are very nearly constant on large scales. For \radc\!, this even extends up to $k\sim$ a few or beyond, depending on halo mass. This indicates that the removal of mass from this region to great approximation determines the change in clustering of these haloes, irrespective of any changes in halo profiles, and that scale-dependent changes in halo clustering (such as seen in the top panels) are simply a consequence of combining different halo mass bins. As we will see in \S\ref{dm_rad}, this remains true when keeping the SO radius fixed. This suggests that one could accurately predict the total power suppression due to feedback by only knowing the fraction of mass that was removed from certain clustered regions. Based on this, we develop a model to do just that in \S\ref{modelmain}.

%$\subsubsection{Constant SO radius between simulations}
\subsection{Keeping the SO radius fixed between simulations}
\label{dm_rad}
To get a sense of the how much matter is ejected from which radius by AGN feedback and where it is deposited, not only do the haloes need to be matched, but a constant halo boundary (SO radius) needs to be taken between simulations. 
This way, the comparison between simulations can be taken one step further, as not only processes within (or outside of) the same haloes, but also within certain regions of space in these haloes can be considered. 

\begin{figure}
\includegraphics[width=1.03\columnwidth]{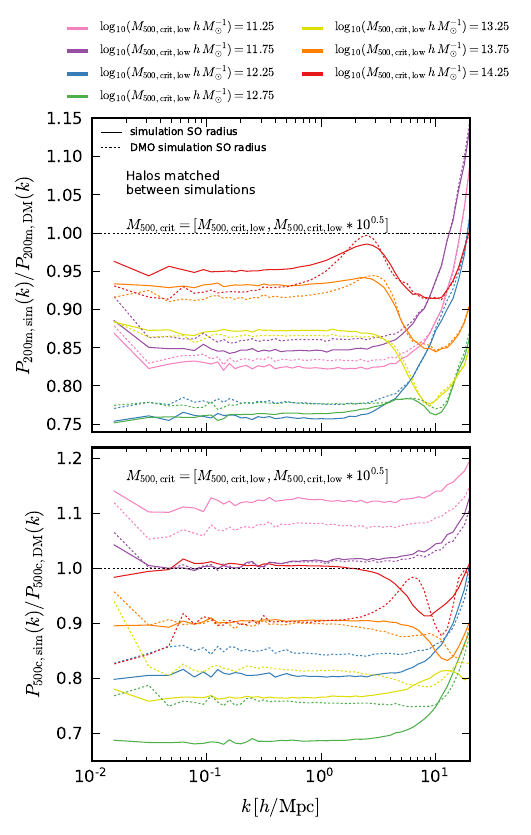}
\vspace{-0.7cm}
\caption{Similar to the bottom-right panel of Figure~\ref{cos_mcm_m_so}, with results for \radm shown in the top panel and those for \radc in the bottom panel, but now with two definitions for the SO radii:
%The dimensionless power spectra (left column) and mass contribution, fraction of full power spectrum of corresponding simulation, (right column) of $C\_AGN\_W7\_L400N1024$ (solid) and  $C\_DMO\_W7\_L400N1024$ (transparent), at $z=0$.
%haloes are matched between DMO and the other simulations. Only matched pairs are shown.
%The legend shows which \M selection is made: $M_{500,crit} \geq 10^{x- 0.25}$ (top row) and $M_{500,crit}=10^{x \pm 0.25}$ (bottom row); $x$ ranging from 11 to 15 (see legend for colour scheme). Masses are measured in C\_DMO\_W7.
either the particle selection criterion is as before, that is using the SO radius measured in each simulation (continuous linestyle), or the SO radius as measured in C\_DMO\_W7 is used for C\_AGN\_W7 as well (dashed linestyle).
The change in SO radius between C\_AGN\_W7 and C\_DMO\_W7 varies between \radm and \radc and with halo mass, with the \radc result being most strongly affected.
}
\label{cos_mcm_mdmr_200}
\end{figure}

In Figure~\ref{cos_mcm_mdmr_200}, the C\_AGN\_W7/C\_DMO\_W7 power ratio is displayed for the usual range in halo masses, for a particle selection where either the simulation SO radius (\radm\!, continuous lines), or C\_DMO\_W7 SO radius (dashed lines) marks the halo boundary.

Although there are slight changes in the ratio between C\_AGN\_W7 and C\_DMO\_W7 power spectra for the \radm simulation radii and the C\_DMO\_W7 \radm radii, the C\_AGN\_W7/C\_DMO\_W7 power ratios of the two regions mostly follow the same trends and have the same shapes (top panel). 
%The top panel shows that for massive mass haloes (e.g. \M=\hM{14.5\pm0.25}, red lines), there is sharp peak in the C\_AGN\_W7/C\_DMO\_W7 ratio at \ksc{\approx}{2} for the DMO SO radius, where the simulation SO radius gives a slow rise with a slightly lower maximum. 
On large scales, a comparison of the lines tells us that there is less mass in C\_AGN\_W7 haloes within the C\_DMO\_W7 \radm radius than in the C\_AGN\_W7 SO radius for massive haloes (\M$\gtrsim$\hM{13.25}), indicating that the C\_DMO\_W7 SO radius is smaller.
The C\_AGN\_W7 \radm has increased compared to the C\_DMO\_W7 \radm because mass has been ejected from inside this radius to outside it, causing the density profile of the haloes to flatten in C\_AGN\_W7.

For lower-mass haloes (\M$\leq$\hM{13.25}), the higher C\_AGN\_W7/C\_DMO\_W7 ratios for the C\_DMO\_W7 SO radius at \ksc{\sim}{0.02} indicate that there is more mass in the C\_DMO\_W7 \radm than in the C\_AGN\_W7 \radm in C\_AGN\_W7, implying a larger C\_DMO\_W7 \radm\!.

The bottom panel of Figure~\ref{cos_mcm_mdmr_200} shows the same as the top panel, except that the halo boundary SO radii are the C\_DMO\_W7 and simulation \radc\!.
%This gives an insight as to what happens to the inner regions of haloes as a consequence of AGN feedback.
At this radius, the consequences of changing the radius to the C\_DMO\_W7 SO radius have much more impact.
Interestingly enough, only haloes with $10^{12.25}\!\lesssim$\M$\lesssim\,$\hM{13.75} (blue, green and yellow lines) have an increased C\_AGN\_W7/C\_DMO\_W7 power ratio for the C\_DMO\_W7 \radc\!, when compared to the simulation \radc power ratio, on large scales. 
The nearest mass bins in both directions (purple and orange lines) suggest no change in C\_AGN\_W7/C\_DMO\_W7 ratio on large scales and the lowest (\hM{11.5\pm0.25}, pink) and highest (\hM{14.5\pm0.25}, red) mass haloes have lower C\_AGN\_W7/C\_DMO\_W7 ratios when choosing the C\_DMO\_W7 \radc instead of the simulation \radc\!.
This suggests that for intermediate mass haloes, the C\_DMO\_W7 \radc is larger than the C\_AGN\_W7 \radc\!; for the more massive or lower-mass haloes, the C\_DMO\_W7 \radc is smaller than or equal to the C\_AGN\_W7 \radc\!.
Therefore, the intermediate mass haloes are dominated by mass ejection inside \radc\!, while the lower-mass and more massive haloes are dominated by contraction.

% A comparison between the top and the bottom panels of Figure~\ref{cos_mcm_mdmr_200} reveals that haloes with \M=\hM{14.5\pm0.25} in particular (red) lose a much larger fraction of mass from inside \radc than inside \radm. 
% On large scales, the C\_AGN\_W7/C\_DMO\_W7 ratio for the C\_DMO\_W7 \radc is comparable to the minimum of the dip for the simulation \radc C\_AGN\_W7/C\_DMO\_W7 ratio, at \ksc{\sim}{8}.
% Therefore, figure \ref{cos_mcm_mdmr_200} shows that most of the mass ejected from \radc by AGN feedback in \hM{14.5\pm0.25} haloes is deposited between \radc and \radm, around \ksc{\sim}{6} from the center (bump in C\_DMO\_W7 \radc line).
% haloes with \M=\hM{14\pm0.25} (orange) show no change in \radc between C\_DMO\_W7 and C\_AGN\_W7, likely due to an interplay between the AGN feedback ejecting matter and the contraction of the halo due to the dark matter reaction to baryon cooling, which peaks the central density.
% haloes with \M=\hM{13.5\pm0.25} (yellow) have a slightly decreased \radm and an increased \radc in C\_DMO\_W7 compared to C\_AGN\_W7.
% Apparently, within \radc, the contraction overrules the impact of feedback and the density profile is peaked, decreasing \radc in C\_AGN\_W7.
% For haloes with \M$\leq$\hM{13.25}, the contraction dominates and is stronger in more massive haloes.
% For the lowest masses, the C\_AGN\_W7/C\_DMO\_W7 ratio is $\geq1$ within \radc, indicating either a strong contraction or that there are too few particles within \radc to obtain reliable results.

The most significant change in the bottom panel is seen for the most massive haloes, where the choice of the simulation or DMO \radc radius not only shifts the contribution by $10\%$ vertically, but also shows either a large bump or large dip for \ksc{\approx}{7}, depending. This suggests that the DMO \radc radius is both smaller than the C\_AGN\_W7 halo radius for these haloes, and simultaneously such that it captures a lot of the material pushed out by feedback. The 
\radc radius in the baryonic simulation, on the other hand, is large enough that the effect of this density bump is effectively smoothed out. To a lesser extent, this effect is also present in the top panel, where the height of the bump at \ksc{\approx}{3} is increased for the two most massive halo mass bins when the DMO \radm radius is used.

\begin{figure*}
    \begin{center}
\includegraphics[width=1.0\textwidth]{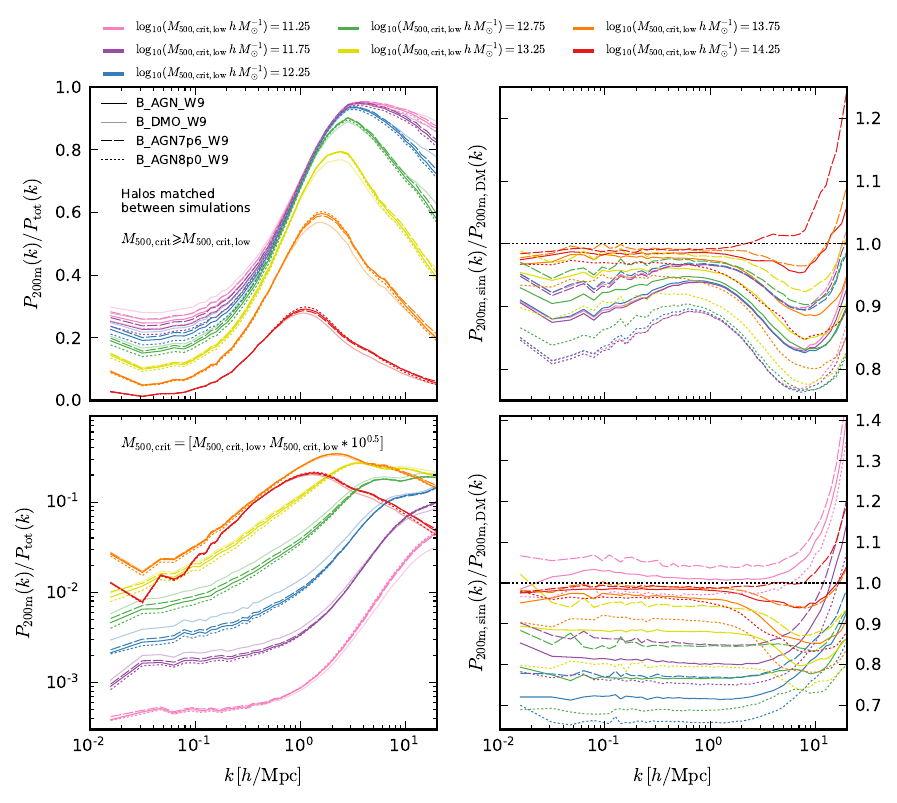}
\vspace{-0.7cm}
\caption{As figure \ref{cos_mcm_m_200}, but showing results for BAHAMAS (B\_AGN\_W9, B\_DMO\_W9, B\_AGN7p6\_W9 and B\_AGN8p0\_W9) rather than cosmo-OWLS.
These simulations shown the change in impact of AGN feedback on haloes of different masses when the strength of the AGN feedback is varied (through the heating temperature). 
The heating temperature has a large influence on the C\_AGN\_W7/C\_DMO\_W7 ratios (right side), while the contribution of the halo mass bins to their corresponding full spectra remain similar.
}
\label{bah_mcm_m_200}
    \end{center}
\end{figure*}

%\subsubsection{Feedback Temperature}
\subsection{Varying the feedback temperature}
\label{fb_temp}
In order to examine the role of the feedback strength on the halo contributions for matched haloes, we can repeat our analysis for the three BAHAMAS simulations where the AGN heating temperature is varied. 
Figure \ref{bah_mcm_m_200} shows the contribution of haloes to the full power spectrum (left side) and baryon-dark matter power ratios (right side) for matched sets of B\_DMO\_W9, B\_AGN\_W9, B\_AGN7p6\_W9 and B\_AGN8p0\_W9, for the same range of halo mass bins as the previous figures. The AGN feedback temperature has a relatively large impact on the B\_AGN\_W9/B\_DMO\_W9 full power spectrum ratio for \ksc{\gtrsim}{0.1}. 
However, the changes in the halo mass bin power spectra and the full matter spectra are of similar order, resulting in a relatively small impact of the AGN feedback temperature on the contribution of each halo mass bin to the full power spectrum as can be seen in the top-left and bottom-left panels of Figure~\ref{bah_mcm_m_200}. The two sets of dashed lines show simulation with a weaker (long-dashed) or a stronger (short-dashed) heating temperature, and show only minor differences with respect to the fiducial feedback strength (solid continuous line).

On large scales, the discrepancies between the contributions to the full power spectrum of B\_AGN\_W9, B\_AGN7p6\_W9 and B\_AGN8p0\_W9 are more substantial around \M$\approx$\hM{12.5-13} (bottom-left panel), suggesting that the masses of these haloes are most sensitive to changes in AGN feedback temperature (as confirmed by the bottom-right panel).

The top-right panel of Figure~\ref{bah_mcm_m_200} indicates that the effect of AGN feedback temperature on the power ratios between the AGN simulations and B\_DMO\_W9 is quite substantial. 
Although it is mostly a vertical displacement, where a higher feedback temperature results in a lower AGN/B\_DMO\_W9 ratio, there are some changes in AGN/B\_DMO\_W9 ratio shape as well. 
For the most massive haloes (\M$\geq$\hM{14.25}, red line), the top and bottom right panels show a clear change in AGN/B\_DMO\_W9 ratio slope at \ksc{\sim}{4}, indicating a change in profile. 
The B\_AGN7p6\_W9/B\_DMO\_W9 power ratio (low AGN feedback temperature) mimics the shapes of the C\_AGN\_W7/C\_DMO\_W7 power ratio, in the top right panel of Figure~\ref{cos_mcm_m_200}. This behaviour indicates that these haloes are hardly affected by the AGN feedback: it is damped so aggressively by their deep potential wells, the effects on the haloes are no longer dominating the shape of the B\_AGN7p6\_W9/B\_DMO\_W9 power ratio.
%The B\_AGN7p6\_W9/B\_DMO\_W9 ratio is instead dominated by the contraction of the halo as it reacts to the presence of baryons in the center.
The B\_AGN8p0\_W9/B\_DMO\_W9 ratio (high AGN feedback temperature) for the most massive haloes (\M$\geq$\hM{14.25}, red line) follows the slope of the other, less massive haloes. Apparently, at this temperature the potential wells of the haloes are no longer deep enough to produce a significant damping effect.
The \hM{14\pm0.25} (orange lines) haloes exhibit a similar, albeit a reduced, effect, as illustrated by the top and bottom right panels.

%Comparing the top-right panel of Figure~\ref{bah_mcm_m_200} to the power ratios between AGN simulations and B\_DMO\_W9 when regarding only particles within \radc (not shown here), generally the same trends are visible.
%However, the vertical offsets between the different simulations are much larger for particles within \radc, especially for massive haloes (\M$\geq$\hM{13.75}).
%The difference between the B\_AGN8p0\_W9/DMO and B\_AGN\_W9/DMO ratio is $\approx0.01$ and $\approx0.04$ for particles within \radm and $\approx0.06$ and $\approx0.1$ for particles within \radc for haloes with \M$\geq$\hM{14.25} (red) and \M$\geq$\hM{13.75} (orange), respectively, at \ksc{=}{0.02}.
%This is a clear indicator that the power suppression in massive haloes caused by AGN feedback is not only stronger within \radc (see Figure~\ref{cos_mcm_m_so}), but also more sensitive to the temperature of AGN feedback within \radc.
Finally, looking at the bottom-right panel, we see the same trends as before: increasing/decreasing the feedback temperature mainly lowers/raises the mass and therefore the relative auto-power contribution at each halo mass, where the largest dependence on feedback temperature is seen around $\sim$\hM{13}. For the most massive haloes, an effect on the halo profile can also be seen.
\\

\subsection{Other contributions to the power spectrum}
\label{othercontributions}
%
%[\emph{Bridge from auto-power to cross-power spectra}]
%
%Apart from the auto-power (halo-halo), there are other contributions to the full power spectrum.
%To produce the whole power spectrum, one would need to add the auto-power of haloes, the auto-power of non-halo particles and twice the cross-power between halo and non-halo particles.
%The previous sections have shown that the power in haloes accounts for most of the power on small scales.
%However, the source of a significant, $>1\%$, fraction of the full power is still undetermined.
%It is either given by cross-power between halo and non-halo particles or by power in non-halo particles.
%As the power spectrum needs to be predicted within $1\%$ accuracy (see \S\ref{intro}), these contributions to the power spectrum need to be taken into account.
%Using the knowledge gained in previous sections concerning the auto-power in haloes, the following section applies the main principles to the cross-power between haloes and all matter to construct a model that captures the effects of baryonic suppression of matter clustering.
%
We have thus far only considered auto-power spectra -- that is, the clustering of haloes of a given mass relative to their own population. While the auto-power spectrum of all halo particles above a given mass (the relative contribution of which was typically shown in the top panels of the preceding figures) provides the dominant contribution to the total matter power spectrum for $k\gtrsim 0.5\,h\,\mathrm{Mpc}^{-1}$, other significant contributions exist, in the form of the auto-power spectrum of non-halo particles (i.e.\ those that live outside of any particular overdensity region) and the cross-power spectrum of non-halo and halo particles. If we choose to separate the haloes into different mass bins, the number of auto- and cross-power terms contributing to the matter power spectrum sharply increases, as we need to add cross spectra of particles in haloes of mass A with those in haloes of mass B (etc), and cross spectra with non-halo particles for all mass bins. It is therefore more convenient (and useful) to instead consider the matter power spectrum as a sum of the cross spectra of non-overlapping groups of particles -- namely, haloes in some particular mass bin, and non-halo particles -- with all matter.

\begin{figure}
    \centering
    \includegraphics[width=1.0\columnwidth]{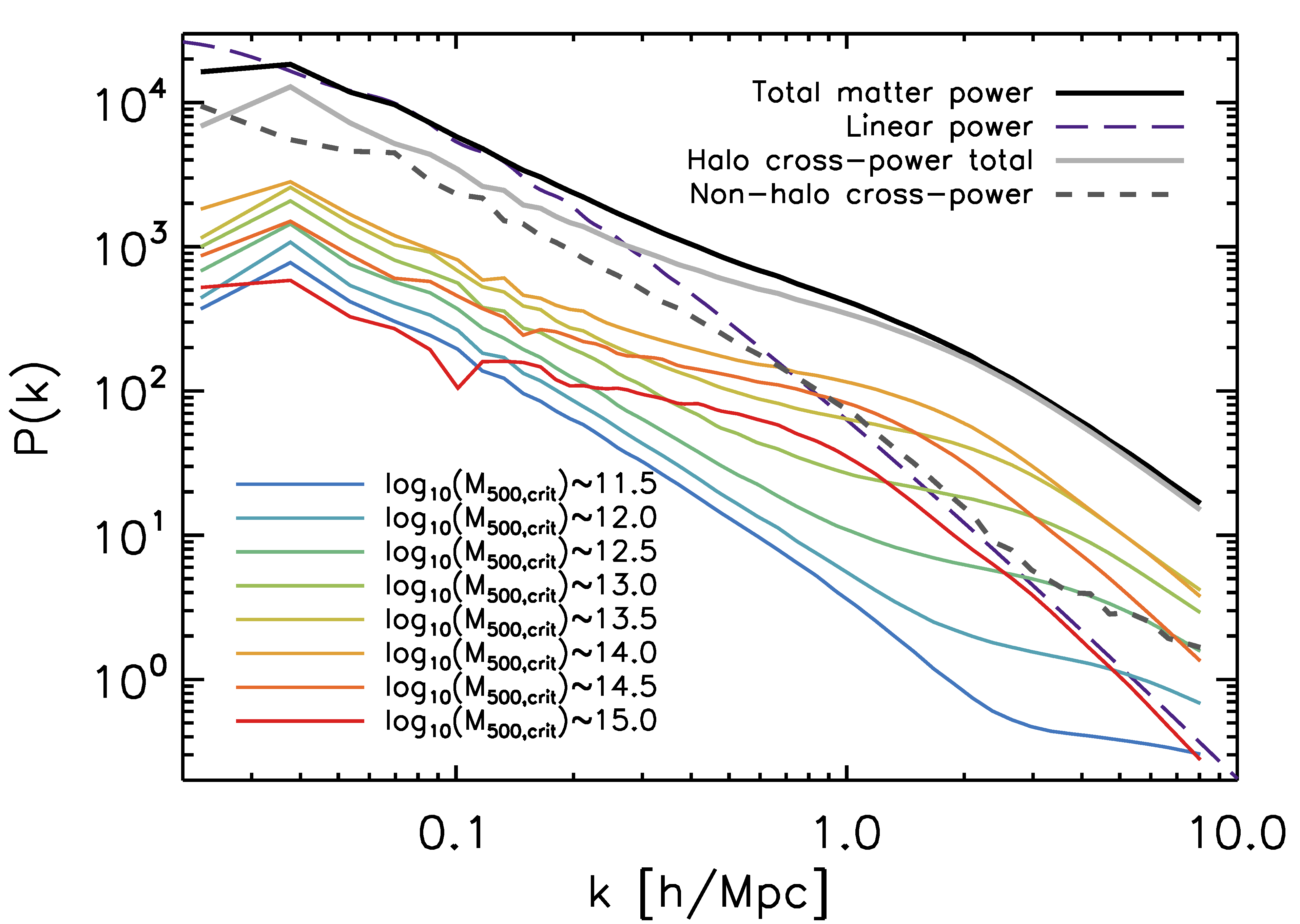}
    \caption{The matter power spectrum of B\_DMO\_W9 (solid black) and its halo (solid grey) and non-halo (dashed grey) cross-power components, $\Pmhtwo$ and $\Pmnhtwo$ respectively (see \S\ref{modelmain}). The mass-weighted halo power in different mass bins, $\fMitwo\Pmhitwo$, is shown as solid colored lines. All haloes are selected on their \M\, mass. A linear power spectrum (long-dashed purple) is shown for comparison. The halo cross-power dominates on small scales, but on large scales halo and non-halo particles contribute roughly equally to the total power, as the total mass is both components is similar.}
    \label{fig:power_components}
\end{figure}

In Figure~\ref{fig:power_components}, we show these cross-power terms for B\_DMO\_W9. To better separate halo from non-halo matter, here we again define haloes as \radm overdensity regions, but still select on their $M_\mathrm{500,crit}$ mass as before. The halo--matter cross terms are shown as coloured lines for different mass bins, with their total shown as solid gray. Adding the non-halo--matter cross term (where non-halo matter is defined here as all matter not in haloes $M_\mathrm{500,crit}>$\hM{11.25}), shown as dashed grey, yields exactly the total matter power spectrum (black). On large scales, the total halo and non-halo cross terms contribute almost equally to the total matter power spectrum, as they contain a comparable amount of mass. On small scales however, the halo cross term fully dominates, as expected. It should be noted that the division into halo and non-halo matter is a function of resolution, as with increased resolution we expect more diffuse matter to be resolved into low-mass haloes -- however, with the current resolution we are already in a regime where the non-halo--matter cross term is subdominant on all scales.

In the following section, we show how we can exploit this division into cross terms to accurately model the power suppression in baryonic simulations.

\section{A ``resummation'' model for the baryonic suppression of matter clustering}
\label{modelmain}
Let us call the cross-power of haloes in mass bin $i$ and all matter, $\Pmhi$(k). Here we will define haloes as regions with an overdensity $\Delta$ relative to the critical density $\rho_\mathrm{crit}$, e.g.\ $\Delta=200\Omega_\mathrm{m}$ or $\Delta=500$. Each halo mass bin will contain some fraction of the total mass in a volume, let us call these fractions $\fM$.

We will refer to the matter-matter auto-power spectrum as $\Pmm$(k). The halo cross terms are assumed to be unnormalized with respect to mass -- the large-scale (linear) halo bias of haloes in mass bin $i$, then, is given by 
\begin{equation}
\label{eq:bias}
\bM=\lim_{k\rightarrow 0}\frac{\Pmhi(k)}{\Pmm(k)}.
\end{equation}
In practice, we will calculate the bias by averaging the power ratio on large scales ($k\lesssim 0.13\,h/\mathrm{Mpc}$), where it is roughly constant.

As discussed in \S\ref{othercontributions}, we can write the matter auto-power spectrum as a sum over all halo-matter cross terms, plus a cross term between all matter and the matter \emph{not} in haloes. This necessarily includes matter in unresolved haloes. Since haloes are highly biased, this last cross term is expected to be dominated by matter just outside haloes (see also \S\ref{fb_temp}). We will refer to the cross term of all matter with matter not in (resolved) haloes as $P_{\mathrm{mnh},\Delta}(k)$ (note that this term, too, depends on our choice of overdensity region). By definition, then, these cross-power terms must satisfy:
\begin{equation}
\label{eq:powersum}
\Pmm(k)=\Pmnh(k)+\sum_i \fM\Pmhi(k),
\end{equation}
where the sum is over all halo mass bins, and thus gives the total contribution of matter in (resolved) haloes.\footnote{Note that for consistency we could also have defined the matter--non-halo cross term as $(1-\sum_i \fM)\Pmnh(k)$, but for the sake of brevity we have absorbed the mass fraction of non-halo matter into $\Pmnh(k)$.}

All quantities considered here can be readily calculated from dark matter only simulations. Figure~\ref{fig:power_components} shows these quantities for B\_DMO\_W9.

\subsection{Accounting for mass loss}
\label{modelmassloss}
Let us now consider what happens to these different terms as matter is redistributed by the processes associated with galaxy formation. While these processes change halo profiles -- for example by gas cooling to small scales and forming stars, contracting the inner dark matter halo -- the main effect on large scales is caused by removing mass from clustered regions. Therefore, on sufficiently large scales we can approximate the effects of galaxy formation by scaling the mass fractions of haloes by the mass they retained -- that is, the ratio of the mass of a feedback-affected halo and its DMO equivalent.

Let us call the mean retained mass fraction for haloes in mass bin $i$, $\fret$. Let us further assume, just for the moment, that the total matter distribution we cross-correlate with is fixed to the DMO distribution. The total contribution to the matter-matter power spectrum of feedback-affected haloes then becomes:
\begin{eqnarray}
\nonumber
\Pmhp(k)\!\!\!&\equiv&\!\!\!\sum_i \fret\fM\Pmhi(k)\\
\label{eq:halopowerprime}
\!\!\!&\equiv&\!\!\!\sum_i \fMp\Pmhi(k),
\end{eqnarray}
where the prime indicates a correction for retained mass.

The mass removed from haloes has to go somewhere, in such a way that the total mass in the volume is conserved. One option is to add this mass to a linear power component -- however, the ejected mass is expected to stay around haloes, and therefore still cluster more significantly than linear. We thus add the mass to the non-halo component, which likely was already dominated by mass around haloes. One wrinkle is that we are removing mass from biased regions -- we therefore need to take bias into account to ensure that the power at low $k$ after modelling galaxy formation still converges to the original value. The ratio of the corrected non-halo cross-contribution to the original non-halo cross-contribution by definition satisfies:
\begin{eqnarray}
\label{eq:nonhalopower_k}
\frac{\Pmnhp(k)}{\Pmnh(k)}\!\!\!\!&=&\!\!\!\!\frac{\Pmmp(k)-\sum_i \fret\fM\Pmhi(k)}{\Pmm(k)-\sum_i \fM\Pmhi(k)}\\
\nonumber
\!\!\!\!&=&\!\!\!\!\frac{\Pmmp(k)/\Pmm(k)-\sum_i \fMp\Pmhi(k)/\Pmm(k)}{1-\sum_i \fM\Pmhi(k)/\Pmm(k)}.
\end{eqnarray}
If we now consider the low-$k$ limit of this expression, and demand that $\lim_{k\rightarrow 0}\Pmmp(k)/\Pmm(k)=1$, we find:
\begin{equation}
\label{eq:nonhalopower_corr}
\frac{\Pmnhp}{\Pmnh}=\frac{1-\sum_i \fret\fM\bM}{1-\sum_i \fM\bM}.
\end{equation}
We thus see that to preserve the power on the largest scale when redistributing mass, we need to not just conserve mass, but the product of mass and bias.

Finally, we have to drop our temporary assumption that the total matter distribution we cross-correlate with is held fixed. Luckily, at this point we already know how the contributions from both the halo and non-halo matter distributions transform, and therefore how the total matter contribution transforms. Using double primes to indicate a correction for halo mass loss in both matter components that make up the cross power, we find:
\begin{equation}
\label{eq:modelq}
\frac{\Pmmpp}{\Pmm}=\left(\frac{\Pmmp}{\Pmm}\right)^2\equiv q_\Delta^2.
\end{equation}
The doubly-corrected halo-matter cross power term then becomes $\Pmhpp=q_\Delta\Pmhp$, and similarly $\Pmnhpp=q_\Delta\Pmnhp$.

In summary, this model takes dark matter only halo mass fractions, linear biases and (cross-)power spectra, and combines them with the mean fraction of mass retained by haloes that have undergone galaxy formation, relative to their dark matter only equivalent, to predict the change in the total matter power spectrum. This mean fraction of mass retained, $\fret$, can be calculated as the total mass ratio of (matched) haloes in a hydrodynamical simulation and its DMO equivalent. Observationally, $\fret$ can in principle be derived from the mean observed baryon fractions of haloes of a certain mass, $\bar{f}_{\mathrm{b},i,\Delta}$, relative to the cosmic baryon fraction. Specifically, under the assumption that all matter removed from the halo was baryonic matter, one would find the retained fraction to be:
\begin{equation}
\label{eq:fbc}
\fbci\equiv\frac{1-\Omega_\mathrm{b}/\Omega_\mathrm{m}}{1-\fbi},
\end{equation}
where $\fbi=M_{\mathrm{bar},i}/M_{\mathrm{tot},i}$ is the mean baryon fraction measured in haloes in mass bin $i$. This can be straightforwardly derived assuming that the ``original'' baryon fraction of each halo is $\Omega_\mathrm{b}/\Omega_\mathrm{m}$, and that the retained CDM fraction, equal to $\fret/\fbci$, is unity.

However, in reality the dark matter will respond gravitationally to the loss of mass, and additional mass will be lost as the halo relaxes. The amount of additional mass lost as a consequence of relaxation after a baryonic ejection event will be explored in Wolfs \& van Daalen (in prep.). For now, we note that the following linear relation is accurate to $\lesssim 1\%$ on average for haloes $M_{\mathrm{h},i,\mathrm{500c}}\gtrsim 10^{12.5}\,\mathrm{M}_\odot/h$ in all cosmo-OWLS and BAHAMAS simulations explored here, regardless of variations in cosmology or feedback strength:
\begin{equation}
\label{eq:fretained}
%\fret\approx 1-a[1-(\fbci+b)],
\fret\approx a(\fbci-b),
\end{equation}
with $a\approx 1.768$ and $b\approx 0.4206$ for $\Delta=200\Omega_\mathrm{m}$. For $\Delta=500$, the best-fit coefficients are $a\approx 2.311$ and $b\approx 0.5251$, though we note that for this overdensity radius the variance is significantly larger (rms $\approx 2\%$).\footnote{We note that $\frettwo$ can be limited to a maximum value of 1 without changing the results presented here, but we find that $\fretfive$ reaches values a few percent higher than this for high-mass haloes.}
Crucially, this relation links the baryon fraction measured at a \emph{retained} total mass (which is what we observe) to a retained fraction at a DMO total mass -- hence no shifting of mass bins is necessary in applying this relationship to equations~\eqref{eq:powersum}-\eqref{eq:nonhalopower_corr}.

For lower-mass haloes ($M_{\mathrm{h},i,\mathrm{500c}}\leq 10^{12}\,\mathrm{M}_\odot/h$), the baryon fraction is more challenging to measure observationally. After checking that these haloes play a negligible role in setting the power suppression up to $k\approx 10\,h\,\mathrm{Mpc}^{-1}$, we fix the retained fractions for these halo masses to unity.

%\begin{figure}
%    \centering
%    \includegraphics[width=1.0\columnwidth]{Figures/crosspower_model.png} %weird legend with pdf version
%    \caption{The true power suppression of B\_AGN\_W9 (red) compared to what our ``resummation'' model predicts. Combining information from both \radm and \radc gives the best results, although baryon fractions inside \radm alone are enough to reproduce the true suppression down to $k\approx 3\,h\,\mathrm{Mpc}^{-1}$ already. Also shown is the simple relation from \citet{2020vanDaalen} (dashed cyan), which was fit to simulations up to $k=1\,h\,\mathrm{Mpc}^{-1}$. Note that we set the suppression to unity for $k<0.07\,h\,\mathrm{Mpc}^{-1}$ and smooth the result to counter sampling noise.}
%    \label{fig:power_model}
%\end{figure}

\begin{figure*}
    \centering
    \includegraphics[width=1.0\columnwidth]{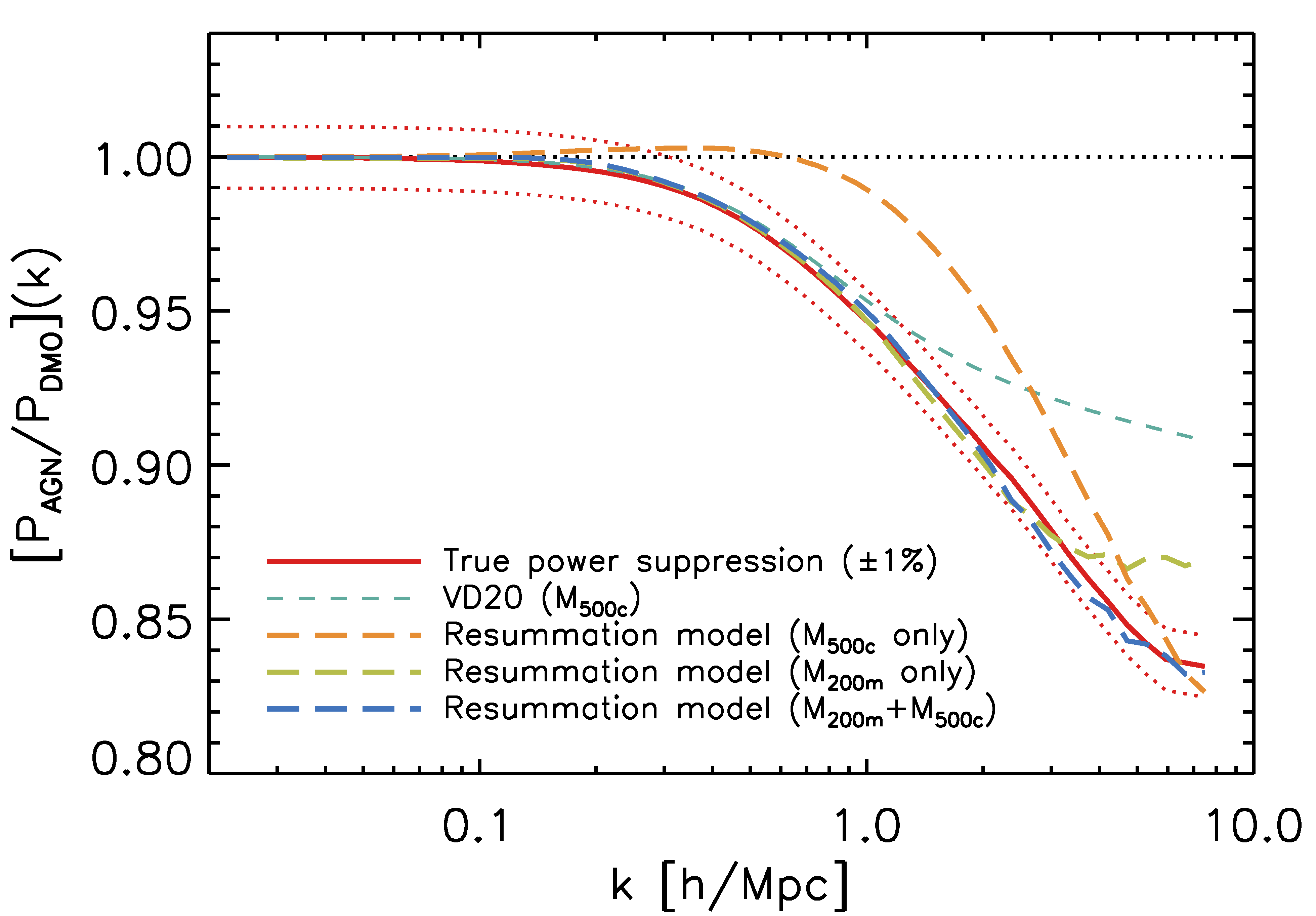} %weird legend with pdf version
    \includegraphics[width=1.0\columnwidth]{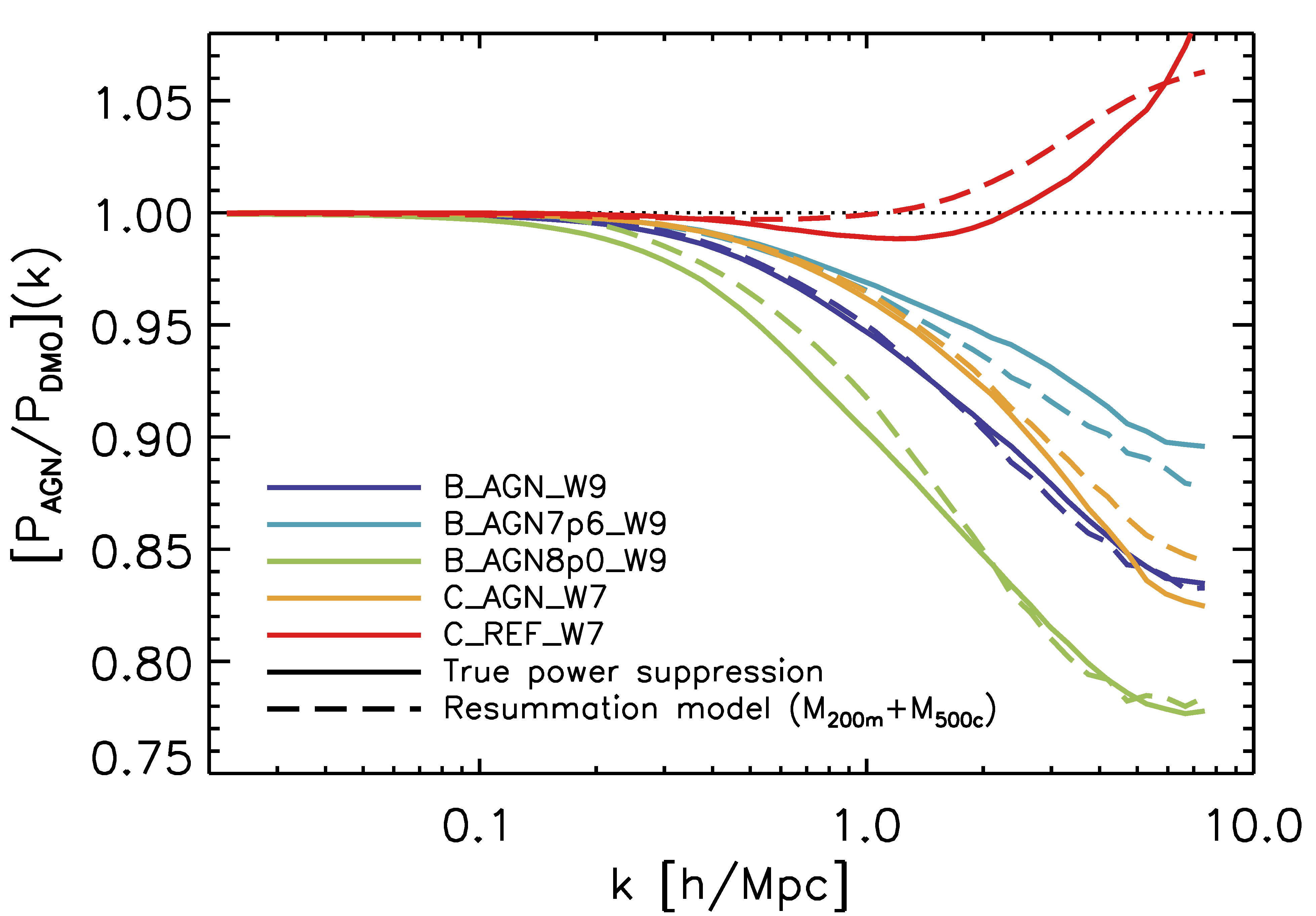}
    \caption{\textit{Left:} The true power suppression of B\_AGN\_W9 (red) compared to what our ``resummation'' model predicts. Combining information from both \radm and \radc gives the best results, although baryon fractions inside \radm alone are enough to reproduce the true suppression down to $k\approx 3\,h\,\mathrm{Mpc}^{-1}$ already. Also shown is the simple relation from \citet{2020vanDaalen} (dashed cyan), which was fit to simulations up to $k=1\,h\,\mathrm{Mpc}^{-1}$ and using only the baryon fraction at a single halo mass. Note that we set the suppression to unity for $k<0.07\,h\,\mathrm{Mpc}^{-1}$ and smooth the result to counter sampling noise. \textit{Right:} The same model applied to all simulations explored here. The ``resummation'' model reproduces the true suppression to well within $1-2\%$ on virtually all scales $k\leq 8\,h\,\mathrm{Mpc}^{-1}$, given baryon fractions as a function of halo mass, using only a single set of two fixed parameter values (see equation~\eqref{eq:fretained}).}
    \label{fig:power_model}
\end{figure*}

\subsection{Combining overdensity regions}
\label{modelannulus}
In principle, we now have all the ingredients to apply this model in practice for a given choice of overdensity region, using the mean baryon fraction measured inside that region for different halo mass bins to rescale and re-sum DMO cross-power spectra. But, if these measurements are available for multiple overdensity regions -- which is now feasible through combinations of X-ray, thermal Sunyaev-Zel'dovich and optical/infrared observations -- we can combine this information to model the power spectra more accurately. As we will show, doing so can greatly compensate for the approximation that the halo profiles are fixed.

Say that we measure the total (retained) mass and baryon fraction of a sample of haloes for both $\Delta=500$ and $\Delta=200\Omega_\mathrm{m}$. We can then calculate the retained fraction for each region with equation~\eqref{eq:fretained}. The net fraction of the ``original'' DMO mass that has been removed from \radc\!, but not \radm\!, is then $\frettwo-\fretfive$. As shown in \S\ref{so_m}, this fraction is quite significant for massive haloes. We will once more refer to the region \radc$<r<\,$\radm as the halo annulus, and will indicate it with subscript $A$. The cross-power contribution of matter in these regions is then given by:
\begin{equation}
\PmAi(k)=\frac{\fMitwo\Pmhitwo(k)-\fMifive\Pmhifive(k)}{\fMitwo-\fMifive},
\end{equation}
and the feedback-corrected annulus cross-contribution is then:
\begin{equation}
\PmAip(k)=[\frettwo\fMitwo-\fretfive\fMifive]\PmAi(k).
\end{equation}
When the annulus contribution can be calculated, we replace the regular $\Delta=200\Omega_\mathrm{m}$ cross-power by $\Pmhitwop=\Pmhifivep+\PmAip$. As before, the fully corrected power contribution is then $\Pmhitwopp=q_\mathrm{200m}\Pmhitwop$.\\

\subsection{Model results}
\label{modelresults}
A proof of concept is presented in Figure~\ref{fig:power_model}. On the left-hand side, we compare the true power suppression measured in B\_AGN\_W9 (red line) to the results of our model (orange, green and blue dashed lines). Our ``resummation'' model takes as input halo baryon fractions measured from B\_AGN\_W9 for haloes \M$>$\hM{12} and combines these with quantities measured from DMO to predict the power suppression due to galaxy formation. Also shown is the result of applying the \citet{2020vanDaalen} relation (cyan short-dashed line), which uses only the baryon fraction measured for haloes around \M$=$\hM{14}. While our model underpredicts the suppression on large scales when only \M\, haloes are used (orange), the agreement between the true suppression and our model's prediction when using only $M_{200,\mathrm{mean}}$ haloes (green) is remarkable. When information from both overdensity regions is combined as described in \S\ref{modelannulus} (blue), the model reproduces the true suppression to $<1\%$ on all scales measured.

On the right-hand side of Figure~\ref{fig:power_model}, we show the results of applying the same model to each of the cosmo-OWLS and BAHAMAS simulations explored in the previous sections, combining the \M\, and $M_{200,\mathrm{mean}}$ overdensity regions. We stress that for each simulation the exact same values for parameters $a$ and $b$ of equation~\eqref{eq:fretained} have been used; the only thing that changes from simulation to simulation are the baryon fractions measured in their haloes (and, when shifting between cosmo-OWLS and BAHAMAS, the DMO simulation used). The model typically reproduces the true power suppression to within $1\%$ accuracy all the way down to at least $k=8\,h\,\mathrm{Mpc}^{-1}$, and almost always within $2\%$. For the model with the strongest AGN feedback, B\_AGN8p0\_W9, the model slightly underpredicts the suppression on large scales, which again hints that haloes in this simulation may lose mass from regions larger than \radm\!. However, even for this case the model is percent-level accurate. Following the procedure outlined in \S\ref{modelannulus}, the model is easily extended to larger or smaller overdensity regions, such as $M_\mathrm{2500,crit}$, to cover an even larger range in $k$ -- as long as baryon fractions can be measured for these regions.

We note that the model is expected to perform slightly worse for cosmo-OWLS, since the DMO simulations of this suite model only a single fluid with a combined dark matter and baryon transfer function, whereas the baryonic simulations separate these into two fluids, and therefore also contain twice as many particles \citep[see Appendix~B of][for more information]{2020vanDaalen}. Despite this, the model and simulations show remarkable agreement for this suite as well.

From the results presented in this work, it is clear why the \citet{2020vanDaalen} relation (shown in cyan in the left-hand panel of Figure~\ref{fig:power_model}) was able to accurately fit the power suppression down to \ksc{\approx}{1} using the baryon fraction of \M$\sim$\hM{14} haloes: (a) the baryon fraction is strongly correlated with the retained mass fraction of haloes; (b) objects around this mass scale dominate the power contributions on large scales; and (c) the effects of feedback in these haloes are generally highly correlated with those in adjacent mass scales. At the same time, this also makes it clear what the limitations of only using a single baryon fraction are: the relation has to implicitly model the link between \hM{14} and other mass scales, capture its scaling with $k$, and be limited to scales roughly comparable to the size of the halo and above. The former two limitations also made the model less universally applicable, as the power suppression in simulations in which the relation between feedback effects on different mass scales deviated strongly from that of other models were more poorly reproduced \citep[e.g.\ the original Illustris simulation, see][]{Vogelsberger2014,2020vanDaalen}. In contrast, while the ``resummation'' model presented here needs additional observationally determined baryon fractions as well as quantities measured from dark matter only simulations as input, it extends to higher $k$ without losing accuracy, could straightforwardly be easily extended to higher redshifts, and has a far less empirical basis.
\\

\noindent
We have employed several implicit assumptions here, such as that feedback does not significantly change the halo positions or radii. The first is justified \citep[see e.g.][]{2014vanDaalen} while the second is not \citep[see e.g.][]{2014Velliscig}, although it is unclear what the impact of this would be, if any. We also have assumed that the linear halo bias is preserved as haloes lose mass. We have checked this assumption by comparing the bias of matched haloes in our simulations, and found that the relative change in bias is typically $\lesssim 1\%$ -- but future iterations of this model could still take bias changes into account to further improve the performance. Another possible improvement is to reduce the scatter in the $f_\mathrm{ret}-f_\mathrm{b}$ relation by taking into account a possible halo mass dependence through the mass-dependent concentration of the haloes. As Elbers et al. (in prep.) will show, taking into account the baryon fraction and concentration simultaneously allows one to predict the power suppression more accurately. Still, from the results presented in Figure~\ref{fig:power_model}, it would appear that our model is already able to yield highly accurate predictions despite these shortcomings.

\section{Summary and discussion}
\label{summary}

In \S \ref{results} we presented our findings concerning the contribution of haloes with a range of masses to the full power spectrum and the varying effects of baryons on the power in these haloes, when compared to a dark matter only simulation, as a function of scale using simulations from the cosmo-OWLS and BAHAMAS simulation projects.
The contributions of various haloes to the full power spectrum and the baryon/DMO simulation power ratios of these haloes were considered for a matched (and an unmatched) set of haloes, their boundary defined by either their \radm or \radc spherical overdensity radius.
These were compared to the results for the halo power within a constant SO radius between simulations.
Lastly, the influences of variations in AGN feedback temperature on the contributions to the full power spectrum and the baryon/DMO simulation power ratios were also investigated for a matched set of haloes.

Our main findings are as follows:

\begin{enumerate}
    \item Haloes of around \M$\sim$\hM{14} provide the dominant contribution to the matter power spectrum up to $k\sim$ a few. Higher-mass haloes provide a lesser but still very significant contribution on large scales (\ksc{\sim}{0.1}), as their rarity becomes more important than their high masses and bias, while lower-mass haloes are less biased on large scales but provide a contribution comparable to $\sim$\hM{14} haloes on smaller scales. This is in line with the findings of \citet{2015vanDaalen}. This also means that the effects of galaxy formation on these haloes are a strong indicator for how galaxy formation affects matter clustering as a whole, as found by \citet{2020vanDaalen}.
    \item By matching haloes between baryonic and dark matter only simulations, we find that galaxy formation including AGN feedback lowers the masses within \radm of haloes down to at least \M$\approx$\hM{11.75}, and therefore also their contributions to the matter power spectrum. On intra-halo scales, the removal of mass from clustered regions by AGN feedback ensures that the contribution of \radm overdensity regions to the matter power spectrum is smaller than that for a DMO universe up to at least \ksc{=}{10}. Including galaxy formation but not AGN feedback also lowers all contributions, but by less than with AGN feedback, and only up to $k\sim$ a few, depending on halo mass.
    \item The halo auto-power for simulations with AGN shows a dip relative to DMO on scales \ksc{\approx}{10} for massive haloes (\M$\gtrsim$\hM{12.75}), corresponding to scales internal to the haloes where mass was removed. For lower-mass haloes, this dip is not present, indicating that these haloes do not host (effective) AGN themselves but were affected by them indirectly. This could be due to nearby AGN heating and driving out the gas from these less massive haloes, or by them affecting the mass accretion history of these haloes.
    \item Even though the auto-power contribution to the matter power spectrum from the same haloes goes down when AGN feedback is included, the peak fractional contribution of haloes $\gtrsim$\hM{13.25} to the matter power spectrum is slightly higher than for DMO. This indicates that the contribution of cross-power between matter inside and outside \radm goes down on scales comparable to this radius.
    \item The choice of halo boundary (\radm or \radc) affects the baryon/DMO power ratios for both massive and lower-mass haloes. Compared to DMO, baryonic haloes with masses $\leq$\hM{12.75} lose a smaller fraction of their mass in their \radc regions than in their \radm regions, and the same goes for the most massive haloes (\M$>$\hM{14.25}). However, haloes with masses $10^{12.75}\leq$\M$\leq$\hM{14.25}, in which AGN are most effective, lose more mass from within \radc than within \radm\!.
    \item By fixing the \radc radius to the DMO value, we find that the auto-power of haloes with $10^{12.25}\!\lesssim$\M$\lesssim\,$\hM{13.75} is increased compared to that within their own \radc\!. The inverse is true for much more or much less massive haloes. This suggests that intermediate mass haloes are dominated by mass ejection inside $\sim$\radc\!, while more massive or lower-mass haloes are dominated by contraction.
    \item The strength of AGN feedback primarily affects the normalization of the halo auto-power contribution as a function of mass, on all scales probed here (\ksc{<}{10}). Only for the most massive haloes, \M$\gtrsim$\hM{14}, does the contribution significantly change shape for $k\gtrsim$ a few, as the AGN feedback affects the halo profiles on these scales.
\end{enumerate}

Furthermore, in \S\ref{modelmain} we presented a novel model that utilizes the cross-power of halo matter inside spherical overdensity regions with all matter in combination with retained mass fractions to predict the suppression of the total matter power spectrum due to galaxy formation. We found that with a simple, fixed relation between observed halo baryon fractions and retained mass fractions our model can reproduce the power suppression of various simulations to $\sim 1\%$ up to at least \ksc{=}{8}. The best results were obtained if both \radm and \radc baryon fractions were available.
\\

\noindent
%Discussion of and future work on the model
Compared to other models for the power suppression, such as halo-model based approaches \citep[e.g.][]{mead2015accurate,debackere2020impact,2021Mead}, baryonification \citep[e.g.][]{SchneiderTeyssier2015,Schneider2020,Arico2021} and models built on fitting the power suppression of hydrodynamical simulations as a function of halo baryon fractions directly \citep[e.g.][]{2020vanDaalen,Delgado2023,Salcido2023}, our ``resummation'' model (\S\ref{modelmain}) has several advantages and disadvantages.

The model presented in this work is built on the link between the observed halo baryon fractions and the retained halo mass, rather than the overall suppression of power, and we find strong indications that this link may not (or only weakly) depend on the strength of feedback, allowing measurements of baryon fractions on different scales to set the power suppression in a model-independent way. There are currently also only two parameters present in our approach, both of which are fixed, and no halo profiles need to be modelled to get accurate predictions up to at least \ksc{=}{8}, which means that cosmological parameters derived from weak lensing observations may have lower uncertainties than in most other approaches, as there are fewer parameters to marginalize over.

The main disadvantage is that our model is built on quantities derived from dark matter simulations, which require significant computational resources -- although many of these already exist in the literature for the purposes of weak lensing analysis. However, similar to e.g.\ baryonification, this shortcoming may be overcome in the future by training a model to produce the halo mass fractions and halo-matter cross-spectra outcomes for such relatively predictable simulations as a function of cosmology, or by using halo model predictions. Exploring the dependence of our approach on cosmology, if any, may also lead to a significant reduction of resources, as fewer dark matter only simulations will be needed if the scaling with cosmology is known. Another disadvantage is that the model requires baryon fractions measured out to \radm\!, which are less readily available than those within \radc\!. While it is possible to predict one from the other, such a mapping would be model-dependent. However, the availability of SZ measurements of $\bar{f}_\mathrm{b,200m}$ is growing, so this may not be an issue for long.

Future work will extend the model presented in \S\ref{modelmain}, including to $z>0$, and apply it to the Gpc-scale FLAMINGO suite of simulations, recently presented in \citet{Schaye2023}. As these simulations contain larger volumes with resolutions similar to BAHAMAS, as well as additional physics variations, analysing these simulations should lead to a more detailed modelling of the relation between the halo baryon and retained fractions, improving the accuracy of the model. Additionally, folding in other overdensity scales such as $r_\mathrm{2500,crit}$ or explicitly including galaxies through the measured stellar fractions of halo may allow us to extend the model to higher values of $k$.

\section*{Acknowledgements and data availability}
We thank the authors of \nbodykit, Nick Hand and Yu Feng, for making their code publicly available. MvD gratefully acknowledges support from the Netherlands Research Council NWO (VENI grant Nr. 639.041.748).

The data underlying this article will be shared on reasonable request to the corresponding author.

%%%%%%%%%%%%%%%%%%%

\bibliographystyle{mn2e}
\typeout{}
\bibliography{Bibliography}

%%%%%%%%%%%%%
\appendix
\section{Box-size effects}
\label{boxsize}

\begin{figure*}
    \begin{center}
\includegraphics[width=1.0\textwidth]{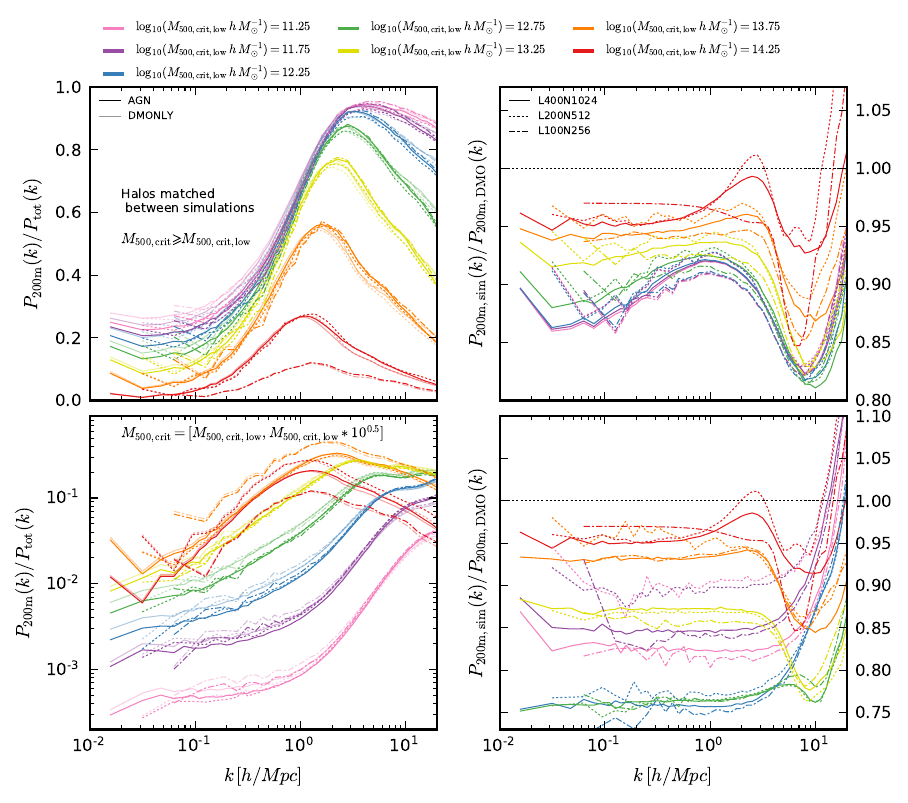}
\vspace{-0.7cm}
\caption{Similar to Figure~\ref{cos_mcm_m_200}, except that not only the L400N1024 (continuous), but also the L200N512 (dashed) and the L100N256 (dot-dashed) cosmo-OWLS simulations are shown. The resolution of all simulations is kept fixed as the box size is varied.
The largest deviations between the different box sizes are in the most massive haloes, due to cosmic variance being large for these in smaller volumes.
}
\label{cos_mcm_m_200_b}
    \end{center}
\end{figure*}

When working with simulation data, there are numerical effects on the data that have no physical cause.
One source of numerical effects is the limited size of the cosmological volume that is simulated.
Apart from the limitation in scales that can be investigated, smaller boxes host fewer haloes disproportionally fewer rare systems, giving larger uncertainties in statistics.

Figure \ref{cos_mcm_m_200_b} shows the contribution of haloes to the full power spectrum and the ratio of AGN/DMO power for a range of halo masses (the same range as used in the rest of the paper), for cosmo-OWLS simulations with volumes of $(400\, h^{-1}\mathrm{Mpc})^3$ (continuous lines), $(200\, h^{-1}\mathrm{Mpc})^3$ (dashed lines) and $(100\, h^{-1}\mathrm{Mpc})^3$ (dot-dashed lines).
The resolution has been kept constant between the simulations.
The haloes have been matched between the DMO and AGN simulations for each box size.
The masses are based on the DMO simulation.

The top-left panel of the figure shows that for most mass bins, the different volumes broadly agree with one another on the contribution of different haloes above a certain mass to the full power spectra.
The minor differences can be attributed to cosmic variance as they show no clear trend with increasing box size.
The clear exception is the contribution of the most massive haloes (\M$\geq$\hM{14.75}) to the full power spectrum in the L100 simulation (red dot-dashed line). 
This is because these haloes are so rare that they have difficulty forming in the L100 simulation and are therefore underrepresented.
The fact that the L200 and L400 simulations converge on the vertical scale shows that these boxes are large enough to give a representational sample.

The bottom-left panel shows convergence for the lower masses (\M$\lesssim$\hM{13.75}) between box sizes.
The two highest mass bins show large discrepancies.
Compensating for the lower contribution to the total of the most massive haloes (\M$=$\hM{14.5\pm0.25}), the second most massive haloes (\M$=$\hM{14\pm0.25}) have a higher contribution to the full power spectrum in the L100 simulation than in L200 or L400.
While the L200 and L400 simulations agree on the contribution of \hM{14\pm0.25} haloes to the full power spectrum, the contribution of \hM{14.5\pm0.25} haloes to the total is higher in L200 than in L400.
As they both agree on the contribution of haloes with \M$\geq$\hM{14.25} (top left panel), it seems a few of the haloes with \M$\geq$\hM{14.75} in the L400 simulation are not fully formed in the L200 simulation, and added to the \hM{14.5\pm0.25} mass bin instead, increasing its contribution to the full power spectrum.

%Upon closer examination, the top left panel shows that the contribution of haloes to the full power spectrum in the L100 simulation is often slightly higher than in the L400 simulation, which is slightly higher than in the L200 simulation, with exception of the highest mass bin.
%Furthermore, the horizontal displacement of the peak for the massive haloes in the L100 and L200 simulations, compared to the L400 simulation, suggests slightly different SO radii.
%Note that while the haloes are matched between AGN and DMO within their boxsize, they are not matched between boxsizes.
%Therefore these slight discrepancies are likely due to differences in the samples.
%Comparison between the boxsizes in the bottom left panel shows that there is no clear trend in the relative positions of the halo power contributions to the total.

The right side of figure \ref{cos_mcm_m_200_b} shows the AGN/DMO power ratios for each halo mass bin for each volume.
The top-right panel shows broad agreement when taking haloes with masses $\gtrsim$\hM{13.5} into account.
As seen in the left-hand panels however, AGN/DMO ratios of the most massive haloes do not agree across the various box sizes.
For haloes with masses $\geq$\hM{13.75} (orange), the depth of the characteristic dip varies with box-size and smaller boxes place the minimum at lower $k$.
Since there is no convergence in the scale at which the minimum of the dip is, larger boxes could still place it at larger $k$ than the L400 simulation.
For haloes with masses $\geq$\hM{14.25} (red), the same can be said about the dip, where the AGN/DMO ratio for the L100 simulation deviates most, due to the under-representation.
Comparing the L200 and L400 AGN/DMO ratios for this mass bin also shows a significant difference at \ksc{\sim}{2-3}, where the small peak in the AGN/DMO ratio is higher (by $\approx0.02$).

The bottom-right panel shows a large discrepancy for the lowest two halo mass bins (\M=\hM{11.5\pm0.25}, pink and \M=\hM{12\pm0.25}, purple) between the L200 simulation and the other two box-sizes for \ksc{\lesssim}{8}.
There seems to be much less mass loss due to AGN feedback in these haloes in the L200 simulation.
Note that these haloes border on the resolution limits.
Apart from the most massive haloes, previously discussed, the AGN/DMO ratios for the other halo masses agree quite well.

As the deviations for the most massive haloes are likely due to the rarity of these haloes, we conclude that a $(400\,h^{-1}\,\mathrm{Mpc})^3$ is a minimum requirement for this study.

%%%%%%%%%%%%%%%%%%%%%%%%

\section{Cosmology}
\label{cosm}

%\begin{figure}
%\includegraphics[width=20pc]{Figures/matched/psdl_WMAP9_L400N1024_032_AD_cosmD.pdf}
%\caption{The power ratios between DMONLY\_2fluid\_WMAP9 and DMONLY\_2fluid\_Planck for the same range of masses as previous figures, at $z=0$.
%The power in \hM{14.5\pm0.25} haloes is lower in the WMAP9 than in Planck simulation.
%For \ksc{\lessim}{2} the power in most other haloes is higher in the WMAP9 simulation (bottom panel). 
%For \ksc{\gtrsim}{2}, the power in the two largest halo mass bins is lower for the WMAP9 simulation, causing the power to be lower in WMAP9 for \ksc{\gtrsim}{2} when haloes with \M$\geq$\hM{13.25} are taken into account (top panel).
%}
%\label{bah_cosmD}
%\end{figure}

Recall from \S \ref{intro}, that the cosmological model parameters influence the power in numerous ways, as they effect halo collapse, formation and evolution, the amount of matter available and the initial power spectrum, resulting, for example, in a different halo mass distribution.
To learn how this translates through into the matter power spectrum and which haloes are most influenced by these changes is crucial, considering the upcoming weak lensing surveys that allow for precision cosmology (see \S \ref{intro}).

\begin{figure*}
    \begin{center}
\includegraphics[width=1.0\textwidth]{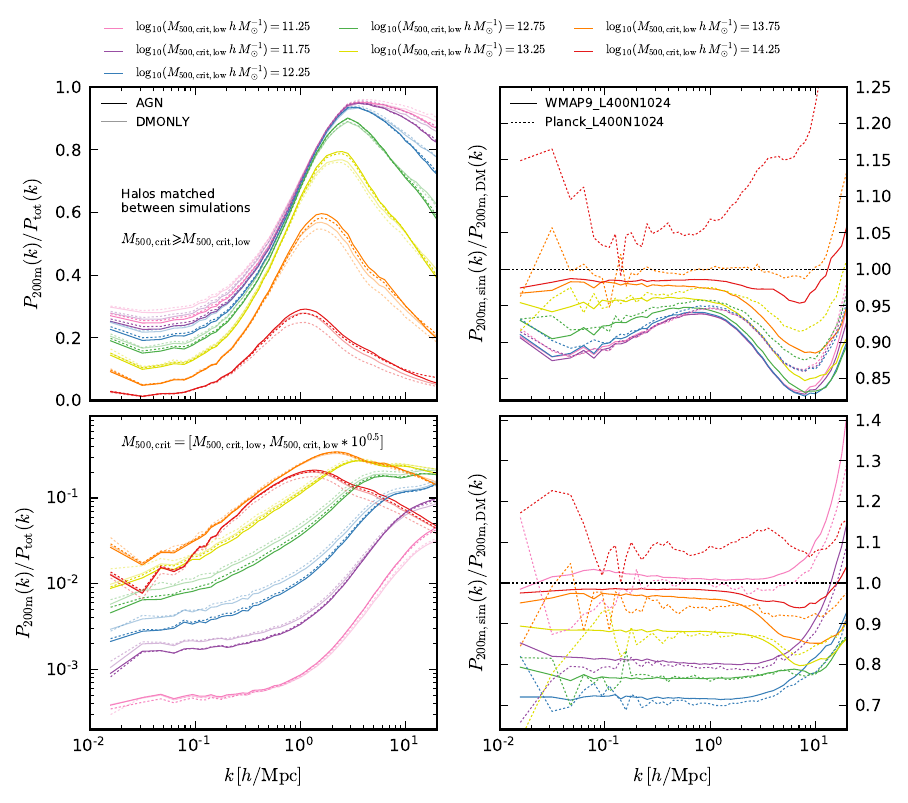}
\vspace{-0.7cm}
\caption{Similar to Figure~\ref{bah_mcm_m_200}, except that the B\_AGN\_PL and B\_DMO\_PL (dashed) have been added to the B\_AGN\_W9 and B\_DMO\_W9 (continuous) power spectra to allow for a comparison between cosmologies.
Both AGN-DMO simulation pairs have been matched and the two DMO simulations have also been matched. The halo mass bins are based on B\_DMO\_W9.
The influence of cosmology on the contributions of halo power to the full power spectrum and the AGN/DMO ratios of halo power is largest for massive haloes (\M$\gtrsim$\hM{13.25}).
Furthermore, the Planck cosmology moves power from massive to lower-mass haloes, when compared to the WMAP9 cosmology.
}
\label{bah_mcm_cosm}
    \end{center}
\end{figure*}

Figure \ref{bah_mcm_cosm} shows that both the contribution of haloes, defined by particles within \radm, to the full power spectrum and the ratio between AGN and DMO for haloes in the usual range of masses are dependent on cosmology, at $z=0$.
For both the WMAP9 (continuous) and the Planck (dashed) simulations the haloes have been matched between AGN and DMO.
Furthermore, the two DMO simulations (B\_DMO\_W9 and B\_DMO\_PL) are also matched, to ensure that every line in figure \ref{bah_mcm_cosm} of a certain colour (mass bin) is based on the same haloes and can thus be compared.
The halo masses as shown in the legend (the same range as in earlier figures) are based on B\_DMO\_W9.

On large scales, the top-left panel shows that the fraction of mass in the massive haloes is similar between the cosmologies.
As more lower-mass haloes are added, the the fraction of mass in haloes in the DMO simulation (transparent) rises above that in the AGN simulations (solid), as in earlier figures, but the two cosmologies broadly agree. 
The Planck cosmology produces slightly more massive haloes for both B\_AGN\_PL and B\_DMO\_PL, due to a higher $\Omega_m$ and a lower $h$, but the top panel in figure \ref{bah_mcm_cosm} shows that the fraction of the total mass in these haloes is also a slightly larger.
This is verified by the bottom-left panel, where the contribution of power in the haloes in the Planck simulation to the full power spectrum is higher for all mass bins in both B\_AGN\_PL and B\_DMO\_PL, with the exception of the most massive haloes (\M$=$\hM{14.5\pm0.25}, red lines) in B\_DMO\_PL and the lowest mass haloes (\M$=$\hM{11.5\pm0.25}, pink lines) in both B\_AGN\_PL and B\_DMO\_PL, the latter having a low matching fraction (see Figure~\ref{match_frac}), thus calling for caution when interpreting.
%Comparing the top left panel of figure \ref{bah_mcm_cosm} to figure \ref{bah_cosmAD}, at \ksc{\sim}{0.02}, it becomes clear that this is because for both the AGN and the DMO simulation, the Planck/WMAP9 halo ratios are higher than the ratios of the full power spectra, even though at this scale, the power in the Planck simulation is lower than the power in the WMAP9 simulation for all halo mass bins shown. 

Moving to the scales where the 1-halo term takes over ($k\gtrsim k_{\mathrm{peak}}$), the top-left panel of figure \ref{bah_mcm_cosm} shows a variation with minimal halo mass in the discrepancies between the two cosmologies in the contribution of halo power to the full power spectrum.
When considering only massive haloes (\M$\gtrsim$\hM{13.5}), the discrepancies between the WMAP9 (continuous) and Planck (dashed) contributions to the full spectrum are quite substantial, where the contributions of the haloes to the full spectrum in the Planck simulations are lower than in the WMAP9 simulations for both DMO and AGN.
Adding lower-mass haloes diminishes these discrepancies, until the roles of the halo power contributions to the total of both cosmologies are reversed and the halo power contributions to the Planck simulations rise above those to the WMAP9 simulations.
This relative position shift of the two cosmologies happens when haloes with masses $\lesssim$\hM{12.75} are added.
%The turnover mass is also the mass around which haloes become massive enough to host an AGN, although this does nothing to explain the behaviour in the DMO simulations.
%As seen in figure \ref{bah_cosmAD}, when adding haloes with masses $\lesssim$\hM{12.75}, the Planck/WMAP9 ratio of the halo power becomes higher than the ratio between the full Planck and WMAP9 power spectra for both AGN and DMO.
This indicates that there is more power in haloes in the Planck simulations and that power that is in massive haloes in the WMAP9 simulation, has been moved to the lower-mass haloes in the Planck simulation. This is likely due to shifts in the halo mass function between these cosmologies.

%At $k$ of a few, the top right panel shows that when taking most masses into account (\M$\geq$\hM{12.75} and lines adding lower masses), the AGN/DMO power ratios match quite closely between the WMAP9 (continuous) and Planck (dashed) simulations.
%Comparing this panel with the Planck/WMAP9 ratios for these mass selections in figure \ref{bah_cosmAD}, we can see the Planck/WMAP9 ratios reflected in the AGN/DMO ratios of the top right panel in figure \ref{bah_mcm_cosm}.
%For \ksc{\gtrsim}{0.7}, the AGN Planck/WMAP9 ratios rise above the DMO Planck/WMAP9 ratios, where they were similar for for lower $k$.
%This is reflected in the AGN/DMO ratios of the top right panel in figure \ref{bah_mcm_cosm}, where the AGN/DMO ratios for the Planck simulation are slightly higher than the AGN/DMO ratios for the WMAP9 simulations, for \ksc{\gtrsim}{0.7}.
%This corroborates the earlier statements that the power in haloes on small scales in Planck is higher than in WMAP9, and that the haloes react stronger to the presence of baryons in the WMAP9 AGN simulation, although the latter statement is still speculation.

For the highest-mass haloes (\M$\geq$\hM{13.25}, yellow, orange and red), for \ksc{\gtrsim}{1}, the differences between the two cosmologies are most striking, as can be seen in the top-left and top-right panels of Figure~\ref{bah_mcm_cosm}.
Next to the previously discussed vertical displacement between the two cosmologies in the top left panel there is also a vertical displacement in the AGN/DMO power ratios.
Moreover, the \hM{14\pm0.25} haloes (orange lines) show the characteristic dip in the AGN/DMO ratio in the WMAP9 simulation around \ksc{\sim}{8} but remain nearly constant for the Planck simulation (bottom right panel).
Comparing the \M$\geq$\hM{13.75} Planck (orange dashed line) and the \M$\geq$\hM{14.25} WMAP9 (red continuous line) AGN/DMO power ratios in the top right panel of figure \ref{bah_mcm_cosm}, reveals these haloes to both omit the characteristic dip around \ksc{\sim}{8}, which shows the reaction of the haloes to the AGN feedback.
The AGN/DMO ratios of these two mass bins are very similar to one another.
This suggests that the potential wells of these haloes in the Planck simulation are so deep, that they resemble the \hM{14.5\pm0.25} haloes in the WMAP9 simulation.
In the Planck simulation the \hM{14\pm0.25} haloes have grown more massive than in the WMAP9 simulation (due to the higher $\Omega_m$ and lower $h$), ending up to be around the size of the \hM{14.5\pm0.25} haloes in WMAP9.
For these group-sized haloes (\hM{14\pm0.25}), cosmology is a very important factor in their formation, growth and structure.
This is also visible, albeit to a lesser degree, in the bottom-right panel of Figure~\ref{bah_mcm_cosm}.

The most massive haloes have an AGN/DMO power ratio $>1$ in the Planck simulation, which is unlikely to be a consequence of bad matching alone, as the matched fraction for these haloes is $>0.95$. This ratio therefore seems to suggest that these haloes have grown so massive that baryonic feedback does not suppress the power at all.
However, even this small deviation from unity in the matched fraction means that there may be a misrepresentation of haloes, which can for example be due to mergers of very massive haloes where the matching has chosen only one of the two haloes, and omitted the other from the set.
As very massive (\M$\geq$\hM{14.25}) haloes are not abundant ($\sim10^2$), this can impact the results substantially.
%Therefore, no conclusive physical results can be drawn from this AGN/DMO power ratio.
%seems to suggest that either because they are more massive in AGN than in DMO for the Planck cosmology.

The bottom-right panel indicates that the DMO/AGN ratios of haloes with masses $\lesssim$\hM{13.25} are only slightly influenced by cosmology, whereas the massive haloes have a much stronger dependence.

In conclusion, cosmology clearly impacts the power contribution of haloes and their AGN/DMO power ratios.
The influence of cosmology is much more substantial for the most massive haloes, in their formation and evolution as well as in the influence of AGN feedback on their power suppression in the AGN simulations, when compared to the DMO simulations.
As there are many factors that are influenced by cosmology that determine the power in different mass haloes, more extensive research is required to disentangle the effects of the various cosmological parameters and find the one(s) dominating the changes in power.

%%%%%%%%%%%%%%%%%

\end{document}